%% file: lrg_environ.tex
\documentclass{aastex}
\usepackage{emulateapj5}
\usepackage{lscape}
\usepackage{onecolfloat}   

\newcounter{affil}
\newcommand{\hoggaffil}[2]{%
	\addtocounter{affil}{1}%
	\altaffiltext{\theaffil}{{#2}\label{#1}}}

\newcommand{\mpc}{\ensuremath{{\rm\,Mpc}}}

\newcommand{\hmpc}{\ensuremath{h^{-1}{\rm\,Mpc}}}

\newcommand{\hmpcC}{\ensuremath{h^{-3}{\rm\,Mpc^3}}}

\newcommand{\ihmpcC}{\ensuremath{h^3 {\rm\,Mpc}^{-3}}}

\newcommand{\msun}{\ensuremath{\rm M_\odot}}

\newcommand{\Mpc}{\mpc}

\newcommand{\nbar}{\ensuremath{\bar{n}}}

\newcommand{\Rmax}{{R_{\rm max}}}
\newcommand{\Rmin}{{R_{\rm min}}}
\newcommand{\xiis}{{\xi_{\rm is}}}
\newcommand{\wis}{{w_{\rm is}}}

\newcommand{\beq}{\begin{equation}}
\newcommand{\eeq}{\end{equation}}
\newcommand{\beqa}{\begin{eqnarray}}
\newcommand{\eeqa}{\end{eqnarray}}

\newcommand{\tableskip}{\\[-6pt]}
\newcommand{\singleline}{\tableskip\hline\tableskip}
\newcommand{\doubleline}{\tableskip\hline\hline\tableskip}

\begin{document}
\twocolumn[

\submitted{Draft in progress, \today}
\lefthead{Small-scale Clustering of Luminous Red Galaxies}

\title{The Small-scale Clustering of Luminous Red Galaxies via Cross-Correlation Techniques}
\author{
Daniel J.\ Eisenstein\altaffilmark{\ref{Arizona},\ref{SF}}
Michael Blanton\altaffilmark{\ref{NYU}}, 
Idit Zehavi\altaffilmark{\ref{Arizona}}, 
Neta Bahcall\altaffilmark{\ref{Princeton}},
Jon Brinkmann\altaffilmark{\ref{APO}},
Jon Loveday\altaffilmark{\ref{Sussex}},
Avery Meiksin\altaffilmark{\ref{Edinburgh}},
Don Schneider\altaffilmark{\ref{PSU}}
}

\begin{abstract}
We present the small-scale ($0.2h^{-1}$ to $7h^{-1}$ Mpc)
cross-correlations between 32,000 luminous early-type galaxies and a
reference sample of 16 million normal galaxies from the Sloan Digital
Sky Survey.  Our method allows us to construct the spherically averaged,
real-space cross-correlation function between the spectroscopic LRG sample
and galaxies from the SDSS imaging.  We report the cross-correlation as
a function of scale, luminosity, and redshift.  We find very strong 
luminosity dependences in the clustering amplitudes,
up to a factor of 4 over a factor of 4 in luminosity, and
measure this dependence with high signal-to-noise ratio.  The luminosity
dependence of bias is found to depend on scale, with more variation on
smaller scales.  The clustering as a function of scale is not a power
law, but instead has a dip at $1\hmpc$ and an excess on small scales.
The fraction of red galaxies within the $L^*$ sample surrounding LRGs
is a strong function of scale, as expected.  However, the fraction of
red galaxies evolves in redshift similarly on small and large scales,
suggesting that cluster and field populations are changing in the same
manner.  The results highlight the advantage on small scales of using
cross-correlation methods as a means of avoiding shot noise in samples
of rare galaxies.  
\end{abstract}

]

\hoggaffil{Arizona}{Steward Observatory, University of Arizona,
		933 N. Cherry Ave., Tucson, AZ 85121}
\hoggaffil{NYU}{Center for Cosmology and Particle Physics, 
    Department of Physics, New York University,
    4 Washington Place, New York, NY 10003}
\hoggaffil{Princeton}{Princeton University Observatory, Peyton Hall,
		Princeton, NJ 08544}
\hoggaffil{APO}{Apache Point Observatory,
                P.O. Box 59, Sunspot, NM 88349}
\hoggaffil{Sussex}{Astronomy Centre, 
		University of Sussex, Falmer, Brighton BN1 9QJ, UK}
\hoggaffil{Edinburgh}{Institute for Astronomy,
		University of Edinburgh, Royal Observatory, Blackford Hill, Edinburgh, EH9 3HJ, UK}
\hoggaffil{PSU}{Department of Astronomy and Astrophysics,
                Pennsylvania State University, University Park, PA 16802}
\hoggaffil{SF}{Alfred P.~Sloan Fellow}

\section{Introduction}

Galaxy clustering allows us to study the relation of galaxies to dark
matter through the biased clustering of dark matter halos and to probe
the possibility of environment-dependent processes in the evolution
of galaxies.  With today's large galaxy surveys, we can measure
this clustering at very high signal-to-noise ratio, investigating
the detailed trends of clustering with luminosity, color, morphology, and redshift
\citep{hubble36,zwicky68,davis76,dressler80,postman84,hamilton88,white88,park94,loveday95,guzzo97,benoist98,willmer98,brown00,Car01,norberg01,zehavi02,norberg02,budavari03,madgwick03,Hog03a,Zeh04b}.

Massive galaxies are particularly interesting to study via clustering
analyses because they tend to reside in massive dark matter halos 
\citep{San72,Hoe80,Sch83,Pos95}.  
These halos themselves have strong trends of clustering amplitude versus
mass \citep{Kai84,BBKS,Mo96,She01}, 
implying that one can find significant variations in clustering
whenever the mean halo mass of the subsample galaxies is changed.
Moreover, because massive galaxies are roughly passively evolving
at low redshifts \citep[][but see Drory et al. 2004]{McC01,Fon04,Gla04}, 
we should be able to interpret the redshift evolution 
of clustering.  A fixed set of galaxies cannot be in high-mass halos 
at one redshift and low-mass halos at the next, whereas a set defined
by a more transitory identifier, such as strong star formation, need not
be found in a consistent location from one time to another.

Here, we use a sample of very luminous early-type galaxies from the 
Sloan Digital Sky Survey (SDSS; York et al. 2000) 
to measure the small-scale clustering 
($0.2h^{-1}$ to $7\hmpc$) of the most massive
galaxies.  We do this by use of angular cross-correlations between
32,000 luminous red galaxies (LRG) with spectroscopic redshifts and 
16 million fainter galaxies from the SDSS imaging survey.
The method of \citet{Eis03} constructs spherically
symmetric measurements of the real-space spatial cross-correlation
function in a flexible manner that supports investigations of 
many different potential trends.  We have previously applied
the method to the study of low-redshift galaxies in the SDSS
\citep{Hog03a,Bla03c}.  We focus here on the luminosity,
scale, and color dependence of the clustering of $\sim\!L^*$ 
galaxies surrounding the LRGs.  

The application to LRGs shows the considerable advantages of the
cross-correlation of spectroscopic and imaging data sets for the study of
small-scale clustering of rare populations.  By avoiding auto-correlation,
we remove a large amount of shot noise from the computation at only a
minor cost in signal.  Moreover the flexibility of the method recommends
it to topics involving trends in multi-dimensional parameter spaces.
We expect that the method will find applications in higher redshift
work as well.

As tracers of high-density regions, massive early-type galaxies also offer
a way to study the environmental dependences of nearby fainter galaxies 
via cross-correlation techniques.
We do this by using the colors of the galaxies in the imaging sample
to measure the fraction of red galaxies as a function of redshift and distance from
the LRG.  Of course, there is a strong gradient in red fraction as
a function of distance \citep{hubble31,abell65,oemler74,melnick77,dressler80,postman84}.
The time evolution of the red fraction as a function of environment 
is one of the more studied quantities in extragalactic astrophysics, 
in clusters as the Butcher-Oemler effect \citep[e.g.,][]{But78,But84,Dre83,Lav86} 
and in the field as the faint blue galaxy problem
\citep[e.g.,][]{Bro88,Tys88,Jon91,Lil91,Met91} and the 
evolution of the star formation density \citep[e.g.,][]{Lil96,Mad96}.  
While all agree that
these reddening trends are a window into galaxy evolution, the questions of whether
this evolution is environmentally dependent and what processes underlie
it remain controversial \citep[e.g.,][]{Has98,Bal99,Mar01,deP04}.

The outline of this paper is as follows.
\S2 reviews the methodology employed,
with more detail provided in Appendices A and B.
\S3 describes the samples of spectroscopic and imaging galaxies we use.
We present the results in \S4 and conclude in \S5.
We use a $\Omega_m=0.3$ flat cosmology throughout and quote magnitudes 
with $h=1$.  All magnitudes have been corrected for reddening
using the \citet{Sch98} map.

\section{Cross-correlation methodology}
\label{sec:method}

We will use the angular cross-correlation method described in
\citet{Eis03}.  In this method, one cross-correlates a sample of known
redshift with a sample of unknown redshift.  The first sample will be
spectroscopic galaxies from the SDSS; the second sample will be galaxies
from the SDSS imaging catalogs.  To the extent that gravitational lensing
can be neglected (justified in Appendix B), 
the only physical correlations are at similar redshift,
and so one can use the known redshift of one object of each pair to
translate angles into transverse distances and fluxes into luminosities
\citep{Dav78,Yee87,Phi87,Lil88,Fer91,Vad91,Sau92,Lor94,Lov97}.  
For example, we will use the known redshift to select a fixed luminosity
range for our imaging sample.

It is well-known that the assumption of isotropic clustering permits one
to deproject angular correlations into true spatial correlations
\citep{von08,Fal77,Dav78,Phi85,Sau92,Bau93,Lov97,Dod00,Eis01z}.
The method of \citet{Eis03} performs this deprojection but notes that
the noisy derivative that usually enters can be avoided by integrating
the real-space spatial correlation function (denoted $\xiis$) on {\it spherical} windows:
\beq\label{eq:Deltadef}
\Delta = {1\over V}\int_0^\infty 4\pi r^2dr\;\xiis(r) W(r)
\eeq
where $W$ is our smoothing window and $V = \int_0^\infty 4\pi r^2dr\;W(r)$.
The resulting quantity $\Delta$ is physically useful: it is the 
average overdensity of objects from the imaging catalog in the 
neighborhood (as defined by $W$) of a spectroscopic object.
$\Delta$ of course depends on $W(r)$, which in turn will have 
a characteristic scale $a$.  Hence, $\Delta$ is a function of $a$,
although we suppress the label.

\citet{Eis03} demonstrates that the quantity $\Delta$ can be estimated
simply by a pairwise summation over the two catalogs with a weighting
function $G(R)$, defined by
\beqa
G(R) &=& {1\over R}{dF\over dR} \\
F(R) &=& {2\over \pi}\int_0^R dr\;{r^2 W(r)\over \sqrt{R^2-r^2}}
\eeqa
Explicitly, we have
\beq\label{eq:Deltasum}
\Delta = {1\over N_{sp}}\sum_{j\in\{sp\}} {1\over \phi_0(z_j)V}\sum_{k\in\{im\}} 
G(R_{jk})
\eeq
where the sums are over the spectroscopic and imaging catalogs, respectively.
$N_{sp}$ is the number of spectroscopic objects, $R_{jk}$ is 
the transverse distance between the two objects (using the angle
and the angular diameter distance to the given redshift), and $\phi_0(z_j)$
is the space density of the objects from the imaging catalog
at the given redshift.  This double summation can then be conveniently
split, to define a value of $\Delta$ for each spectroscopic object:
\beq\label{eq:Deltaj}
\Delta_j = {1\over \phi_0(z_j)V}\sum_{k\in\{im\}} G(R_{jk}).
\eeq
We can tabulate these $\Delta_j$ for each spectroscopic object and then 
take the average for any subset we wish.  This allows us to find 
the run of overdensity versus redshift and luminosity and to perform
jackknife resamplings with essentially no overhead.  It also makes
clear that there is no need to bin the spectroscopic galaxies in redshift before computing.
Appendix A describes how we account for masks and efficiently handle the summation
over large-separation pairs.

Note that the space density of the imaging catalog as a function of
redshift, $\phi_0(z)$, is not necessarily known.  By choosing a sliding
flux limit in the imaging catalog as a function of the spectroscopic
object's redshift, we minimize the change in $\phi_0(z)$ while also 
keeping the reference sample as constant as possible, so that clustering
amplitudes can be most easily related to clustering bias.
It is important to note that even if $\phi_0(z)$ is unknown, the
relative values of $\Delta$ between different scales and different
luminosities of the spectroscopic galaxies are still accurately known.
In other words, we can study differential effects at a single redshift
and even the redshift evolution of those differential effects.  Only
the absolute clustering amplitudes as a function of redshift depend
on the modeling of $\phi_0(z)$.

We will use windows of the form
\beq\label{eq:Wr}
W(r) = {r^2\over a^2} \exp\left(-{r^2\over 2a^2}\right)
\eeq
where $a$ is a scale length.  This choice allows us to focus
on about one factor of two of scale in the cross-correlation $\xiis(r)$.
The apodization at small scale improves on a pure Gaussian window
in that it downweights very small-angle pairs where object deblending might
be a concern.

For this choice, the required $F$ and $G$ functions are
\beqa
F(R) &=& 4a^2 s e^{-s} \left[ 2s I_0(s) - (1+2s) I_1(s) \right] \\
G(R) &=& s e^{-s} \left[ (6-8s) I_0(s) + (-2+8s) I_1(s) \right] 
\eeqa
where $s = R^2/4a^2$ and $I_0$ and $I_1$ are modified Bessel functions
of the first kind.  For coding, it is useful to note that
\beq
G(R) \approx {-1\over \sqrt{2\pi s^3}} \left[ {3\over 4} + {45\over 32s} + {1575\over 512 s^2} + O(s^{-3}) \right]
\eeq
for large $s$.  We see that $G \approx 3R^2/2a^2$ at small $R$, so small
separation pairs are indeed downweighted.  For this window, $V = 3 a^3 (2\pi)^{3/2}$.

For a power-law $\xiis(r) \propto r^{-1-\alpha}$,  
the resulting connection between $\Delta$ and $\xiis$ is
\beq\label{eq:xi2Delta}
\Delta = {2\over 3} \sqrt{2\over \pi} \sqrt{2}^{-\alpha} \Gamma\left(2-{\alpha\over2}\right) \xiis(a).
\eeq
For $\alpha=0.8$,
this means that $\Delta = 0.36\xiis(a) \approx \xiis(1.76a)$.
For $\alpha=1.0$, $\Delta = \xiis(a)/3 \approx \xiis(1.73a)$.
In other words, we are roughly measuring the correlation function
on the scale $1.75a$.  This shift is not surprising in that $W(r)$
has most of its weight at $r>a$.

We will compute the mean environment for many different luminosity
bins, redshift bins, scale lengths, and two imaging galaxy samples.
The errors on each data point are estimated by jackknife resampling.
We use 50 spatially coherent subsamples of the data for the resamplings.
We have investigated the covariances between different measurements.
Of course, different scale lengths and imaging samples for the same
primary LRG bins are highly covariant, typically 50-70\% correlated 
between scales separated by a factor of two.  However, we find that different
luminosity and redshift bins are generally close to independent.  This
is expected: there are not many LRGs in a given large non-linear structure.

We will use proper (i.e.~non-comoving) distances throughout this paper.
The reason is that with this choice the computed quantity $\phi_0\Delta$
is predicted to be redshift independent for the stable clustering ansatz \citep{Pee80}.
This is easy to see because $\phi_0V\Delta$ is the number of imaging galaxies
surrounding the spectroscopic galaxy, and stable clustering predicts that
these structures are time independent.
On much larger scales, linear perturbation theory \citep[e.g.][]{Pee80}
predicts $\phi_0\Delta\propto (1+z)^{-n}D^2$ for a fixed proper scale, 
where $D(z)$ is the growth function (perhaps
generalized to include the evolution of bias) and $n$ is
the power-spectrum spectral index $P\propto k^n$; for concordance 
cold dark matter cosmologies, this prediction is also close to constant
if the bias is time-independent.  Aside from the aesthetics of a constant
baseline hypothesis, the choice of non-comoving distances also makes it
simple to average over redshifts if the spectroscopic sample is imperfectly
volume-limited.  In practice, we average $\phi(z)\Delta$ rather than $\Delta$,
as the former is empirically close to constant in redshift.

Appendix B argues that the effects of weak lensing are only of order 1\% 
for the results in this paper.  This is slightly below our best 1--$\sigma$
errors.  We therefore neglect gravitational lensing.  False correlations
imprinted by photometric calibration errors common to the two samples are
shown Appendix A to be negligibly small for this application.

\pagebreak
\section{SDSS samples}\label{sec:sdss}
\subsection{Description of the SDSS}

The SDSS \citep{Yor00,Sto02,Aba03,Aba04} is imaging $10^4$ square degrees away from
the Galactic Plane in 5 passbands, $u$, $g$, $r$, $i$, and $z$
\citep{Fuk96,Gun98}.  Image processing \citep{Lup01,Sto02,Pie02}
and calibration \citep{Hog01,Smi02} allow one to select galaxies,
quasars, and stars for follow-up spectroscopy with twin fiber-fed
double-spectrographs.  The spectra cover 3800\AA\ to 9200\AA\ with
a resolution of 1800.  Targets are assigned to plug plates with a
tiling algorithm that ensures nearly complete samples \citep{Bla01t}.

Galaxy spectroscopic target selection proceeds by two algorithms.
The primary sample \citep{Str02}, referred to here as the MAIN
sample, targets galaxies brighter than $r<17.77$.   The surface
density of such galaxies is about 90 per square degree.  The LRG
algorithm \citep{Eis01} selects $\sim\!12$ additional galaxies
per square degree, using color-magnitude cuts in $g$, $r$, and $i$
to select galaxies to $r<19.5$ that are likely to be luminous
early-types at redshifts up to $\sim\!0.5$.  The selection is
extremely efficient, and the redshift success rate is very high.
A few galaxies (3 per square degree at $z>0.15$) matching the
rest-frame color and luminosity properties of the LRGs are extracted
from the MAIN sample; we refer to this combined set as the LRG
sample.  In detail, there are two sections of the LRG algorithm,
known as Cut I and Cut II and described in \citet{Eis01}.

We begin from a spectroscopic sample covering 3,836 square degrees.  
The exact survey geometry is expressed in terms of spherical polygons 
and is known as {\tt lss\_sample14}.
This set contains 55,000 spectroscopic LRGs in the redshift range $0.15<z<0.55$.

\subsection{The spectroscopic LRG sample}

The SDSS LRG sample is nearly volume-limited, but not perfectly so.
In particular, the flux limit of the survey creates a significant
luminosity threshold at redshifts above 0.37.  We therefore focus
on two volume-limited subsamples in this paper:  $0.20<z<0.36$ with $-23.2<M_g<-21.2$,
and $0.20<z<0.44$ with $-23.2<M_g<-21.8$.  Here, the $M_g$ is the
rest-frame $g$-band absolute magnitude at $z=0.3$.
This is computed from the observed $r$ magnitude using
the $k$ and passive evolution corrections of the ``non-star-forming''
model presented in Appendix B of \citet{Eis01}.
We evolve the galaxies to $z=0.3$ rather than $z=0$, in order to keep the 
results closer to the observations.
See \citet{Zeh04c} for more 
details and plots of the number densities of these samples.
The first sample totals 26,000 LRGs; the second totals 12,500 LRGs, with about 6000 new ones.

Note that because the angular cross-correlation method does not rely
on comparisons of the positions of any two spectroscopic galaxies,
the exact geometry or redshift selection function of the LRG sample
is not required.  However, it is relevant to note that the LRG selection
uses a luminosity cut that depends on rest-frame color.  Intrinsically redder galaxies
can enter the sample at lower luminosities \citep{Eis01}.  This means that our results
at lower luminosities ($M_g\gtrsim-21.5$) are preferentially from the red edge of the red 
sequence of early types.  However, essentially all LRGs are from the
red sequence, which is intrinsically quite narrow \citep{Fab73,Vis77,Bow92}, 
indeed narrower than
the photometric errors in a typical SDSS observation.   
\citet{Hog03a} computed the mean density as a function of luminosity
and color in the MAIN sample, where there is no color selection.
They found no gradient of density with color at the high luminosity end.
Hence, between the photometric errors across the red sequence and the
null result of \citet{Hog03a}, we expect any gradient with respect to 
color to be small.

In this paper, we at times wish to express the $M_g$ magnitudes in terms
of luminosities relative to $L^*$.  We have adopted $M_g^*=-20.35$ for
this purpose.  This number was derived from the value of $M^*=-20.44$
in the $^{0.1}r$ band at $z=0.1$ \citep{Bla03b}, applying an evolution of 0.31 mag to 
$z=0.3$, and converting from the $^{0.1}r$ band to the $g$ band using
a mid-type spectral energy distribution.  We note, however, that this
$M_g^*$ value is only approximate; the $M_g$ magnitudes of the LRGs are
the quantities closer to the observations.

\begin{table}[b]\footnotesize
\caption{\label{tab:magz}}
\begin{center}
{\sc Luminosity Cuts \\}
\begin{tabular}{ccccccc} 
\tableskip\tableline\tableline\tableskip
&&& \multicolumn{2}{c}{``1.0'' Sample$^{c}$} & \multicolumn{2}{c}{``0.4'' Sample$^{c}$} \\ 
$z$ & $m^*_r$$^{a}$ & $\delta(g-i)$$^{b}$ & $\phi(z)$$^{d}$ & Red$^{e}$ & $\phi(z)$$^{d}$ & Red$^{e}$ \\
\tableskip\tableline\tableline\tableskip
\input{environ_tab3.dat}
\tableskip\tableline\tableline\tableskip
\end{tabular}
\end{center}
NOTES.---%
$^{a}$The $r$-band apparent magnitude $m^*$ we adopt as a function of redshift
to approximate a constant set of galaxies.  
The results 
roughly match the value of an $L^*$ galaxy at $z=0.3$ 
in an $\Omega_m=0.3$ flat universe.
However, the model likely slides off of this reference, in the 
sense that at lower redshift we are using more luminous (and
hence less abundant) galaxies.\\
$^{b}$The tolerance in observed $g-i$ to define a ``red'' galaxy.
The color must be within $\delta(g-i)$ of the red sequence at the given redshift.\\
$^{c}$We use two samples, one from $m^*-0.5$ to $m^*+1.0$, the
other $m^*-0.5$ to $m^*+0.4$, denoted as the ``1.0'' and ``0.4'' 
samples, respectively.  \\
$^{d}$$\phi(z)$ is the comoving density for this
magnitude range predicted by taking a representative sample from the
SDSS MAIN sample \protect\citep{Bla02,Bla03b} and moving the galaxies to the given 
redshift.  Magnitude evolution of $Q=1.6$ was assumed.  The numbers
in the table are in units of $10^{-3}\ihmpcC$.  Note, however, that
elsewhere in this paper, we use proper densities and so
$\phi_0  = (1+z)^3\phi(z)$. \\
$^{e}$The fraction of those galaxies predicted to be classified as ``red''.
\end{table}

\pagebreak
\subsection{Galaxies from SDSS Imaging}
\label{sec:imaging}

We use 5305 square degrees of imaging from the SDSS to define our second
sample of galaxies.  The imaging completely covers the spectroscopic sample,
but the extended region is useful in that one can compute correlations
to LRGs that are near the spectroscopic boundaries (which are numerous
because of the plate coverage of the SDSS).  We extract all galaxies 
down to $r=21$.

At each redshift considered, we use only a fraction of galaxies.
Ideally we would select a consistent set of galaxies at all redshifts,
but this is not possible in fine detail.  We have attempted to select a
fixed luminosity range, corrected for evolution.  To do so, we define
a reference galaxy and find its $r$ band magnitude as a function of
redshift in the usual cosmology, then select galaxies within a
specified range relative to this magnitude.

Our reference magnitude, denoted $m^*$, is computed from an early-type
galaxy spectral energy distribution, including $k$ corrections and
passive evolution.
The resulting $r$-band reference magnitudes as a function of redshift
are given in Table \ref{tab:magz}.
We have scaled the model so that it matches our estimate of $L^*$ at $z=0.3$.

We consider two luminosity ranges, $M^*-0.5$ to $M^*+0.4$ and $M^*-0.5$ to $M^*+1.0$.
The former allows us to use galaxies up to $z=0.44$ at $r<21$, while
the latter reaches $z=0.36$ (the approximate volume-limited redshift of the LRG
sample) with more galaxies.  We refer to these as the ``0.4'' and ``1.0''
samples, respectively.

The $m^*$ values are indeed close to that of an $L^*$ galaxy, but not perfectly so.
In particular, our early-type model does not evolve as much as recent
luminosity functions \citep[e.g., $Q=1.6$,][]{Bla03b} and the red color
makes $m^*$ fainter at higher redshift than a typical galaxy would.  As 
a result, we estimate that $m^*$ is actually diverging from the actual
track of an $L^*$ galaxy by about 0.2 magnitudes per 0.2 in redshift,
in the sense that at lower redshift we are using more luminous and less
abundant galaxies.  For typical luminosity functions in this luminosity range,
every 0.1 mag of shift alters the densities by 10\%.  Therefore, we 
expect that our samples may contain anomalous evolution at the level
of $(1+z)^{1.5}$.  
Given the mild trends of mean environment with luminosity around $L^*$ found in 
\citet{Hog03a}, it seems unlikely that the modest luminosity shifts in our
modeling would significantly alter the bias properties of the imaging sample
across our redshift range.

While we surely could construct a model that would match $L^*$ more
carefully, it is not obvious what property of galaxies one should actually
track.  For example, since $\phi_0\Delta$ is the number of selected
galaxies near the LRG, one might be more interested in defining the sample
so that cluster early-types were being consistently selected, which is
likely to be closer to what we have done.  Indeed, as galaxies near and far
from LRGs may evolve differentially, there may be no truly comoving selection.
Our view is instead empirical: Table
\ref{tab:magz} defines the apparent magnitude ranges that we have used,
and redshift-dependent interpretations should work to this definition.

We have taken MAIN sample galaxies from {\tt lss\_sample14} to construct
a luminosity-color function \citep{Bla02,Bla03b} and used that sample to predict
the luminosity function of galaxies in the observed $r$ band at each redshift.
The resulting comoving number densities are in Table \ref{tab:magz}.
We assume luminosity evolution of $Q=1.6$ (1.6 magnitudes per unit redshift),
independent of color, but we include no evolution of the spectral energy
distributions themselves.
As expected, the number density increases by about 20\% per 0.2 in redshift.
Most of this is due to the favorable $k$ correction of blue galaxies 
boosting their number in the $r$-band sample.

We find that the SDSS luminosity function predicts a comoving number
density of about $\phi_{com}=0.01\ihmpcC$ and $\phi_{com}=0.045\ihmpcC$ for the 
``1.0'' and ``0.4''
samples, respectively.
Note, however, we will be using physical distances for our $a$, so one should
use $\phi_0 = \phi_{com}(1+z)^3$ when converting from $\phi_0\Delta$ to $\Delta$
and hence to $\xiis$.

Because LRGs are often in clusters or rich groups, we will be interested in their
correlations with red galaxies as well.  
We use the same $r$-band flux cuts as above and then
define a red galaxy as one
that is within a particular tolerance in $g-i$ color of the empirical color
of the red sequence of early-type galaxies as a function of redshift in
the SDSS.  We set the tolerance in observed $g-i$ to match an 0.08 mag
blueward shift in rest-frame $g-r$ color as a function of redshift; the 
resulting values are listed in Table \ref{tab:magz}.  The considerable 
increase in this observed color range as a function of redshift is 
because the spectral energy distributions of red and not-so-red galaxies
diverge as one moves blueward over the 4000\AA\ break.
Note that our color cut is a mix of evolving and non-evolving models:
the central color is the empirical one and hence is evolving with redshift,
whereas the color range is derived from non-evolving templates.

Table \ref{tab:magz} lists the red fraction,
i.e. the ratio of the number density of the red galaxies to that of all galaxies 
in the $r$-band range,
predicted by our modeling with the SDSS luminosity-color function.
The red fraction is predicted to drop toward higher redshift
despite the fact that there is no evolution of spectral energy distributions
or differential evolution of red and blue galaxies in our modeling.
The primary reason is that the $k$-corrections of the blue galaxies are more
favorable at higher redshift, such that for a given apparent $r$-band
magnitude, one is looking further down the the luminosity function of
blue galaxies than that of red galaxies.  A secondary reason is that in this 
model galaxies are evolving in luminosity somewhat faster than our $m_r^*$
reference value, so that one is selecting slightly lower luminosity and hence
slightly bluer galaxies at higher redshift.  

Despite our effort to match rest-frame colors, it is possible that our 
selection of red galaxies is redshift dependent.  Modeling errors in 
some aspect of the color selection or in the luminosity evolution of 
the galaxy population would
be one cause; another would be the neglect of photometric scatter across
the color boundaries.  
As before, our sample definition is empirically precise, but modelers 
will need to account for our selection.

To conclude, we stress that while uncertainties in the evolution of our samples
must be addressed to interpret bulk redshift trends, the uncertain $\phi_0$
values cancel out and
are no concern when studying trends with scale and LRG luminosity within volume-limited
samples such as are employed here.  In addition,
one can study the redshift evolution of trends with scale and luminosity.

\section{Results}\label{sec:results}
\subsection{Scale and Luminosity Dependences}

\begin{figure}[tb]
\plotone{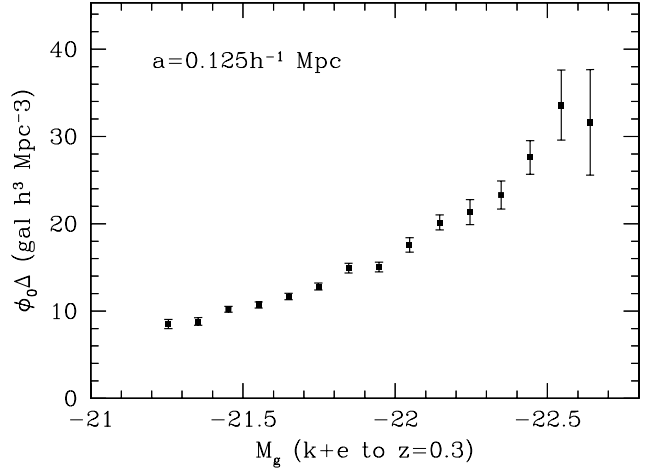}
\caption{\label{fig:a125}%
Mean overdensity ($\phi_0\Delta$) around red galaxies as a function of luminosity,
expressed as the rest-frame $g$ magnitude, $k+e$ corrected to $z=0.3$.
The window is $a=0.125\hmpc$, proper.  Only galaxies in
the range $0.20<z<0.36$ have been used.  The ``1.0'' imaging sample 
is used, so the galaxies are in the range $M^*-0.5$ to $M^*+1.0$.
}
\end{figure}

In Figure \ref{fig:a125}, we present $\phi_0\Delta$ with a proper scale length
of $a=0.125\hmpc$ for LRGs in the redshift range $0.20<z<0.36$ with 
regard to imaging galaxies in the range $M^*-0.5$ to $M^*+1.0$ (the ``1.0'' sample).
The LRGs have been binned into $0.1$ magnitude bins from $M_g=-21.2$
to -22.7 (i.e. $2.2L^*$ to $8.3L^*$).  It is immediately clear that
the density of galaxies around the LRGs is a strong function of
LRG luminosity, ranging from $\phi_0\Delta=8$ to $\sim\!30$.

What do these numbers mean?  $\phi_0$ is the proper number density of the 
imaging sample galaxies, $\sim\!0.022\ihmpcC$ at $z=0.3$.  Hence, 
we have $\Delta$ running from $\sim\!400$ to $\sim\!1500$.  
This is the value of the cross-correlation
function between LRGs and $\sim\!L^*$ galaxies, 
averaged over the window $W(r)$ and centered at $\sim\!0.22\hmpc$.
Note that the classical $\xi(r) = (r/r_0)^{-1.8}$ with $r_0=5\hmpc$ comoving 
would predict $\Delta=170$,
but LRGs are known to be highly biased \citep{white88,norberg01,Zeh04b,Zeh04c}.

Alternatively, if one multiplies $\phi_0\Delta$ by the volume of $W(r)$
(here, $V=3(2\pi a^2)^{1.5} = 0.0922\hmpcC$), then one has the average
number of $M^*-0.5$ to $M^*+1.0$ galaxies (above a random unclustered
floor of $\phi_0V$, insignificant for $a\lesssim0.5\hmpc$ but not for $a\gtrsim1\hmpc$) 
surrounding the LRG,
with the counting weighted by $W(r)$.  This latter interpretation
is independent of $\phi_0$ and so our numbers are quite precise,
e.g. for $\phi_0\Delta = 10.8$, there is 1.0 galaxy, weighting the count by $W(r)$,
in the range $M^*-0.5$ to $M^*+1.0$ around that class of LRGs.
Note that $W(r)<1$, so the weighted count is less than the actual number of galaxies.
We will show $\phi_0\Delta$ when we want to appeal to the correlation
function and $\phi_0V\Delta$ when we want to stress the empirical accuracy.

More approximately, if one considered the density of galaxies to be constant
in the window region, then $\phi_0\Delta$ would be that density (proper,
in this case).  In other words, the proper density of $M^*-0.5$ to $M^*+1.0$ galaxies 
$0.2\hmpc$ from an LRG is about 10--$30\ihmpcC$.  Of course, since the density
is in fact steeply declining with scale, this interpretation is not precise.

\begin{figure*}[p]
\plotone{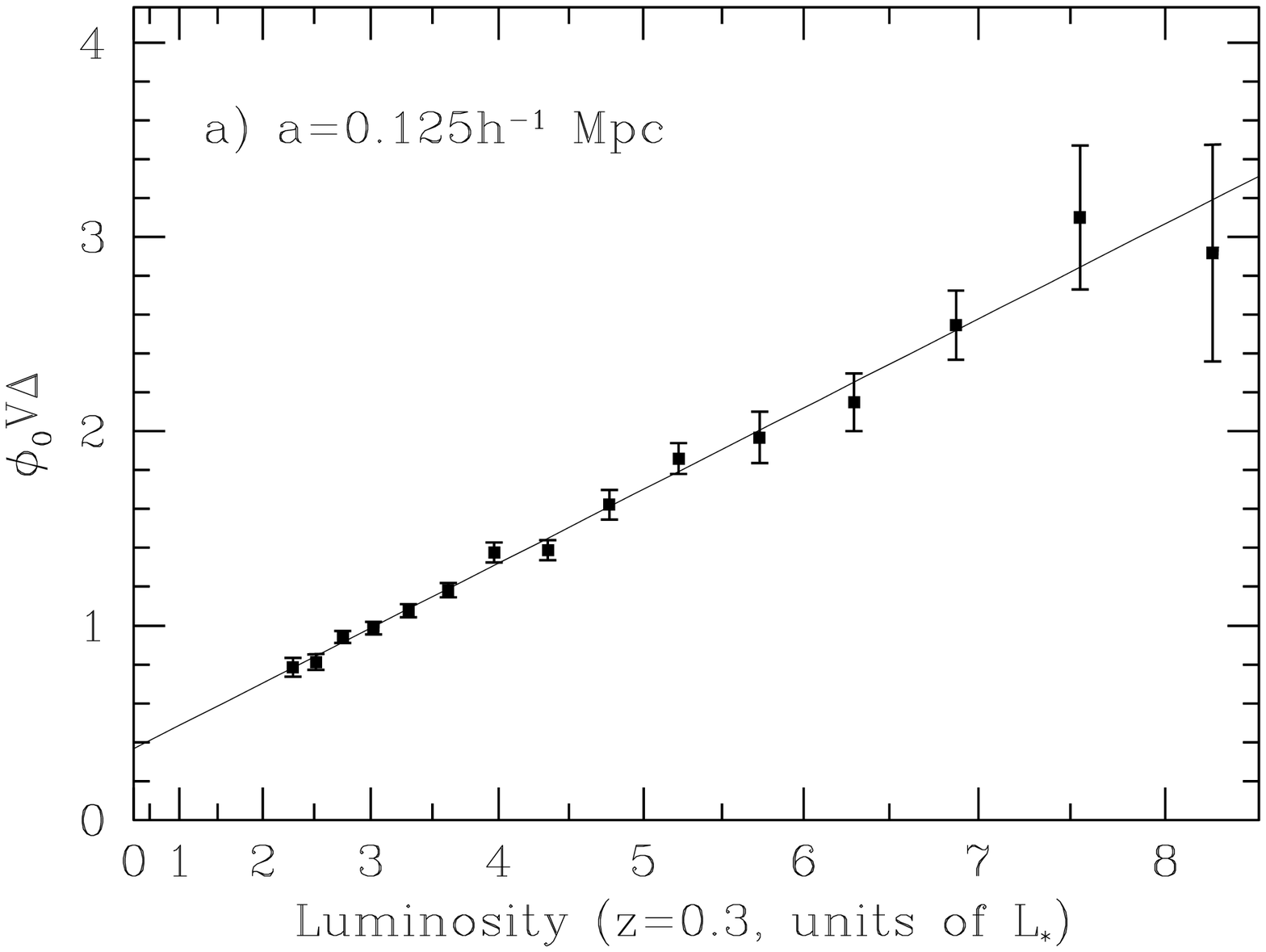}
\plotone{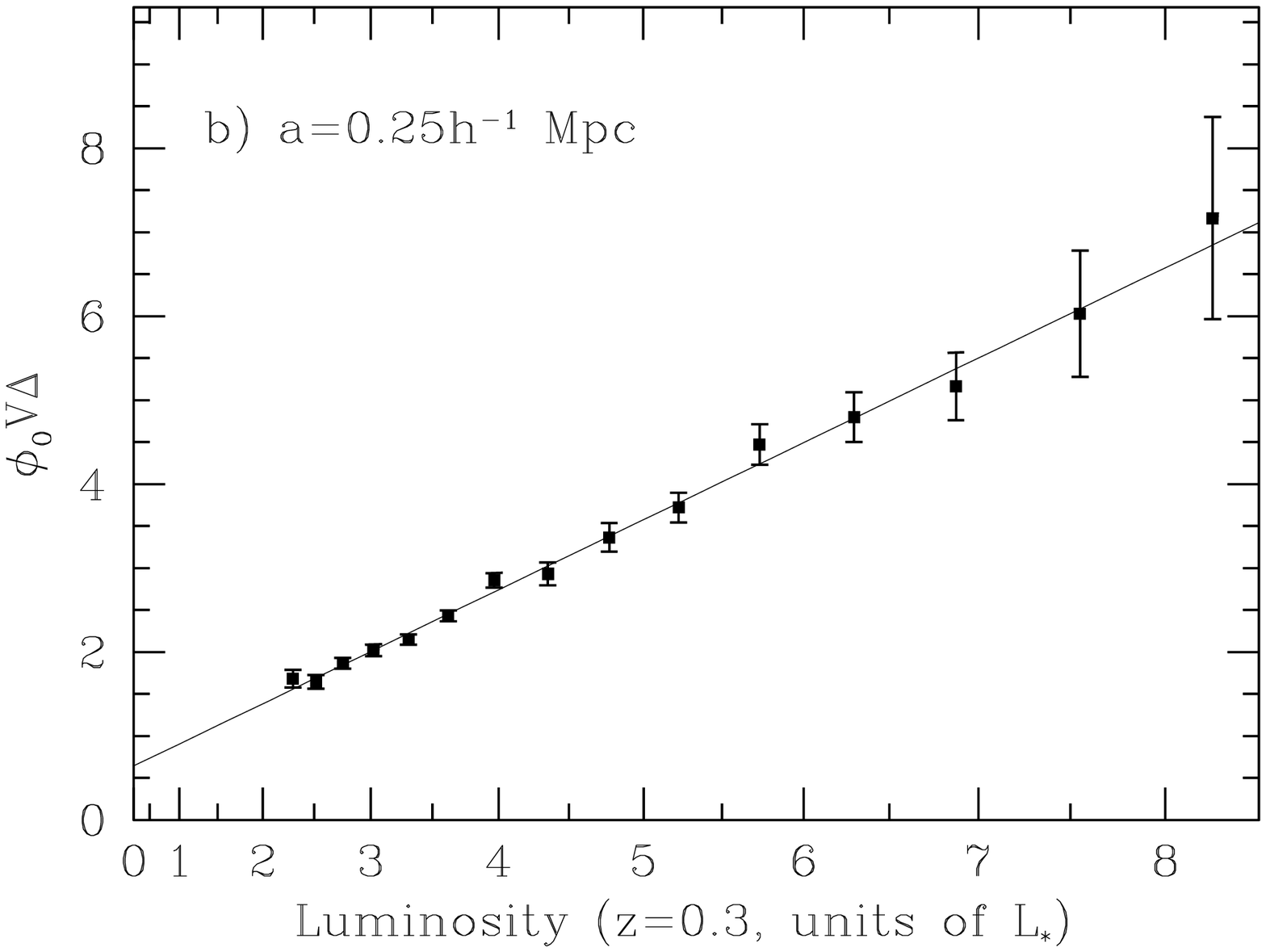}\\
\plotone{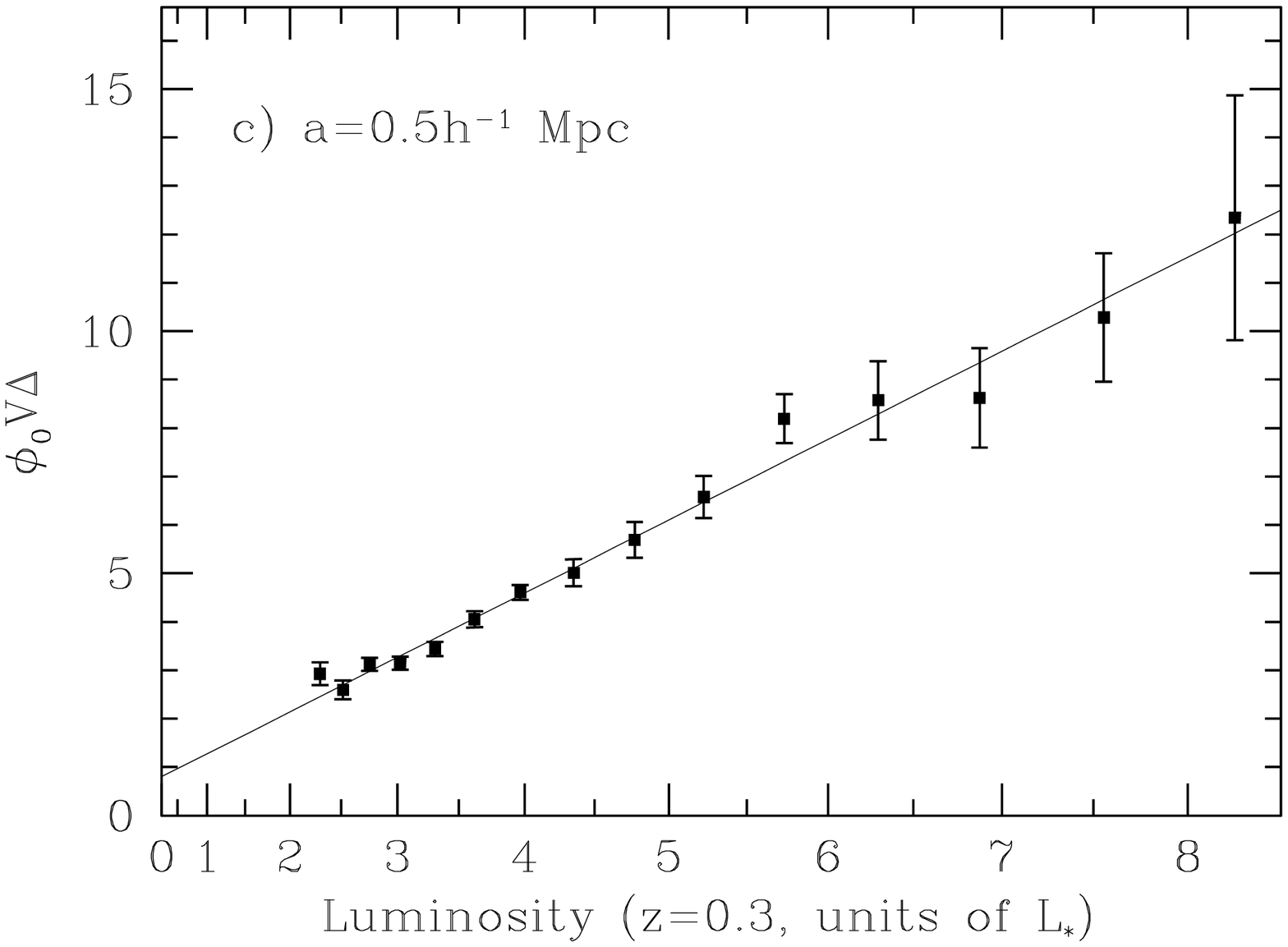}
\plotone{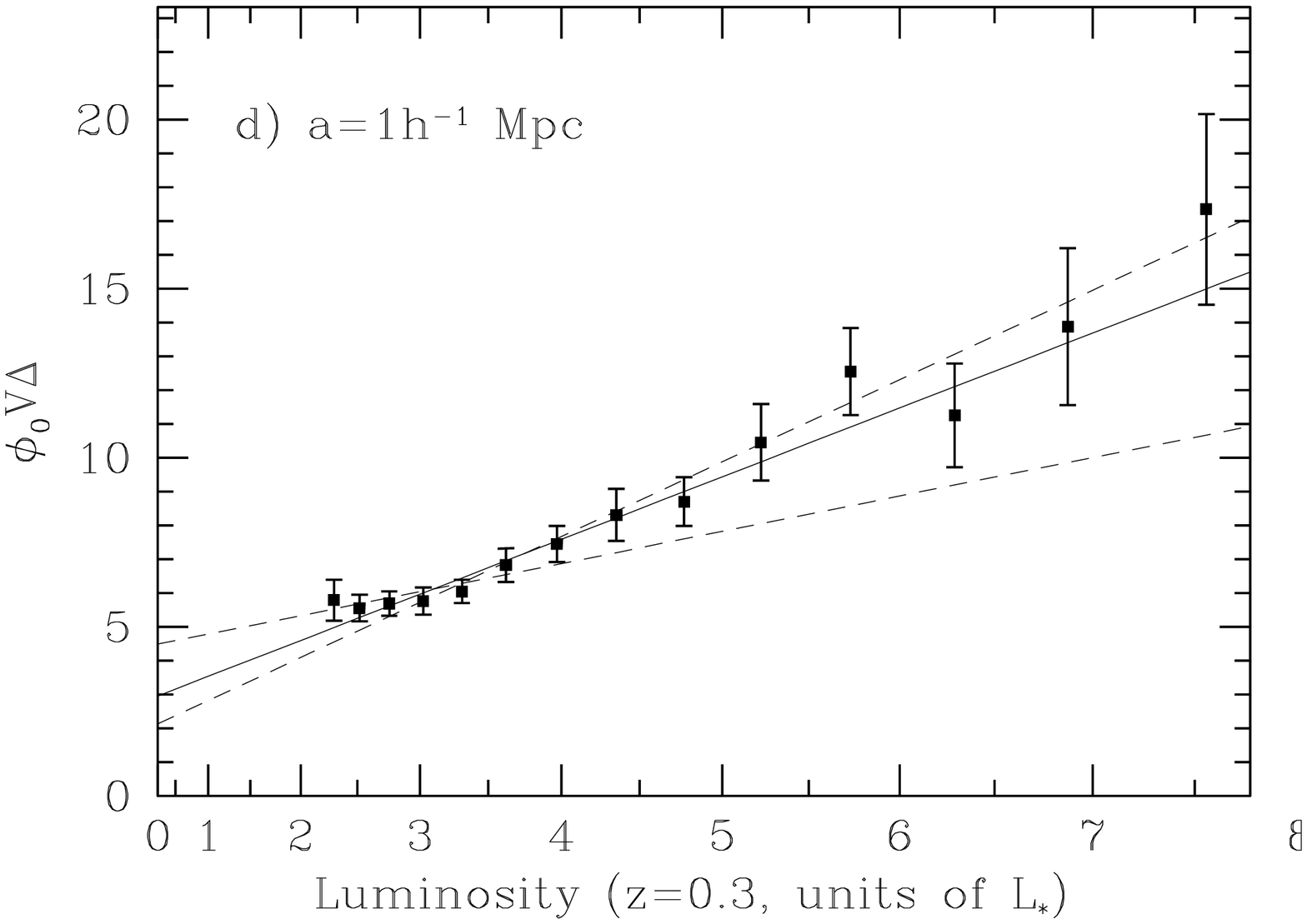}\\
\plotone{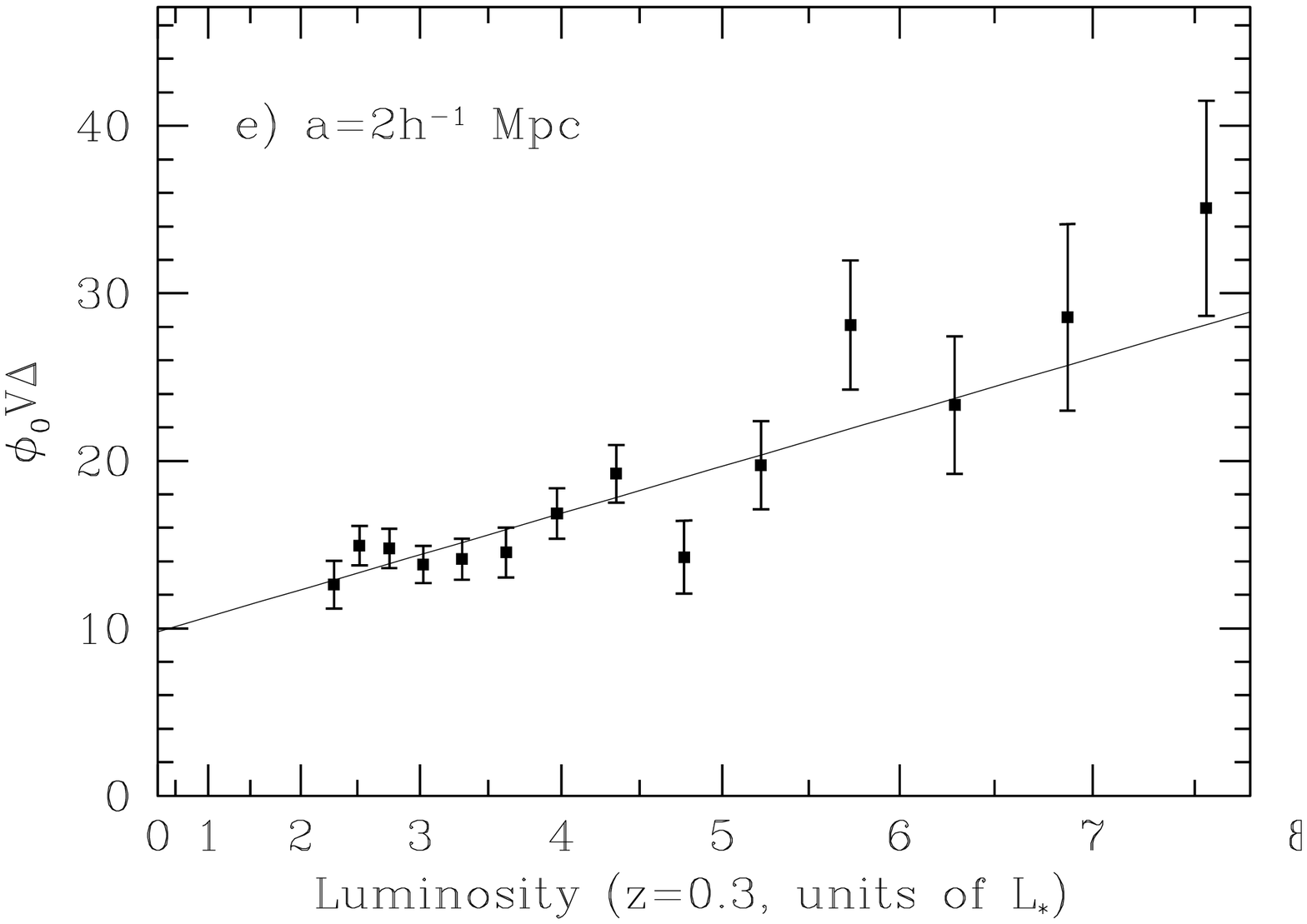}
\plotone{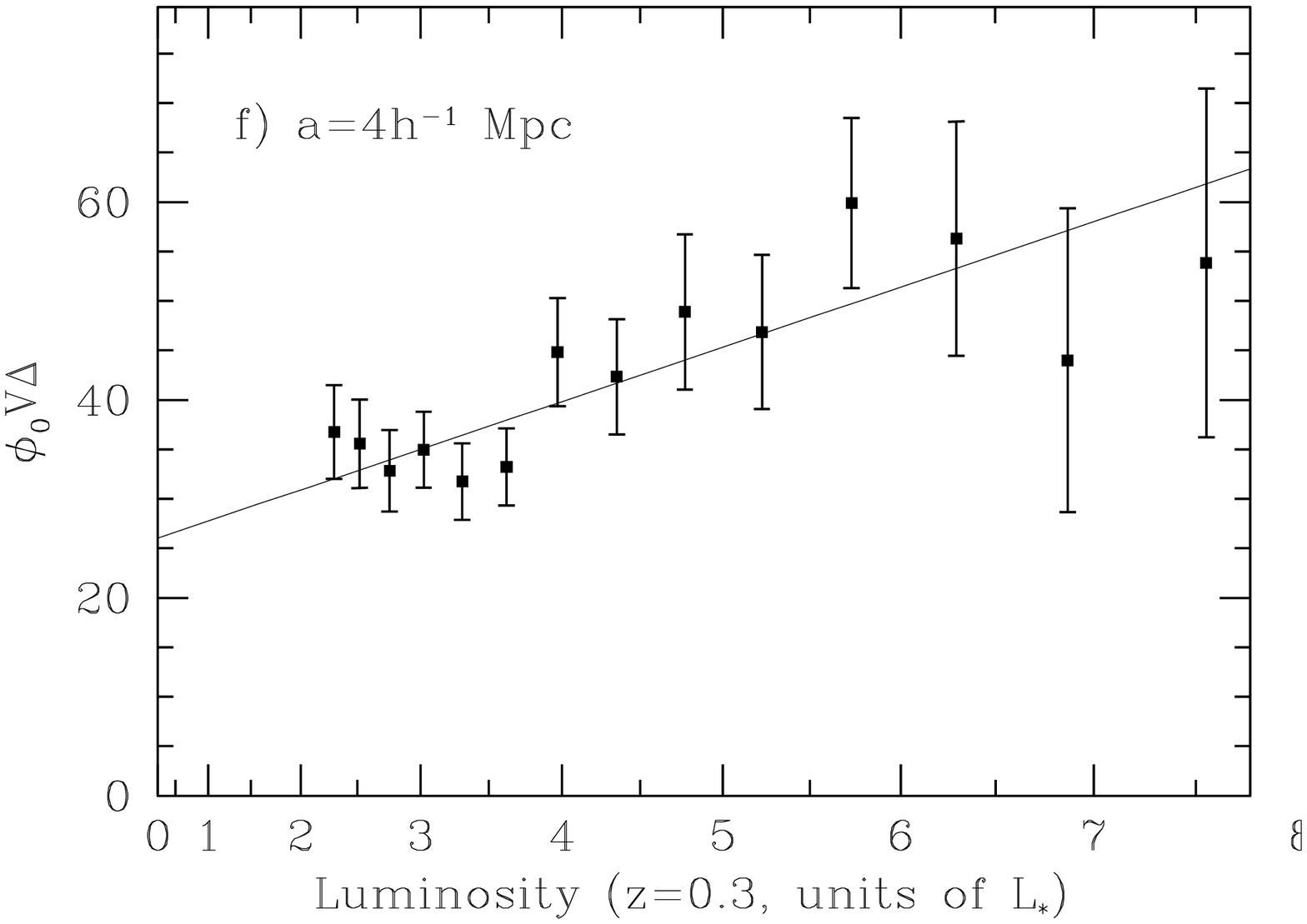}\\
\caption{\label{fig:a6}%
Mean overdensity ($\phi_0V\Delta$) around red galaxies as a function of luminosity,
expressed relative to $M_g^*=-21.35$.  The horizontal axis has
been stretched to scale as luminosity to the 1.5 power, as this
makes the linear fit better in most cases.
The solid lines are the best $AL^{1.5}+B$ fit, reported for these and other samples
in Tables \protect\ref{tab:fits212} and \protect\ref{tab:fits218}.
({\it a-f}) Six panels show different scales: $a=0.125\hmpc$, $a=0.25\hmpc$, $a=0.5\hmpc$,
$a=1\hmpc$, $a=2\hmpc$, and $a=4\hmpc$, proper, respectively.  
The dashed lines in panel (d) are the best-fit lines from $a=0.125\hmpc$
(steeper)
and $a=4\hmpc$ (shallower) rescaled and overplotted to show how the slope is 
changing with scale.
Recall that the $\Delta$ statistic is in fact probing the correlation function
on scales of about $1.75a$, so the range here is roughly $0.2h^{-1}$ to $7\hmpc$.
Only galaxies in the range $0.20<z<0.36$ have been used.  The ``1.0'' imaging sample 
is used ($M^*-0.5$ to $M^*+1.0$).
}
\end{figure*}

One might worry that because $\phi_0$ is a reasonably strong function of $z$
(see Table \ref{tab:magz} and recall the extra factor of $(1+z)^3$), then
deviations in the mean redshift as a function of luminosity would compromise
the comparison of $\Delta$ from Figure \ref{fig:a125}.
In fact, the samples are volume-limited to sufficient accuracy:
the mean redshift of the higher luminosity bins in Figure \ref{fig:a125}
is only 0.004 higher than that of the lower luminosity bins (0.296 vs.\ 0.292), 
which is only 1.2\% more in $(1+z)^4$.  This demonstrates that $\phi_0$ is 
essentially constant across the LRG luminosity range being used.

With the result that $\phi_0$ is the same for all of the luminosity bins, Figure
\ref{fig:a125} shows a four-fold variation in $\Delta$ as a function
of LRG luminosity, which means a four-fold variation in the small-scale
cross-correlation of LRGs with respect to $L^*$ galaxies!  

It is difficult to quote an exact value of the bias of the LRGs with
respect to the mass because of the uncertainty in $\phi_0$ and the
fact that we are using a particular set of galaxies (namely, $L^*$
galaxies, as defined by Table \ref{tab:magz}) to trace the density field.  
Formally, we are measuring
the cross-correlation between LRGs and $L^*$ galaxies.
There is the temptation to interpret this cross-correlation as 
varying as the product of the bias of LRGs with respect to mass
and the bias of $L^*$ galaxies with respect to mass.
Since the latter is constant, one would interpret the change in $\Delta$
as only the change in the bias of LRGs.
However, galaxy bias need not be separable in this simple fashion. 
Although $L^*$ galaxies seem to have bias close to unity as a whole 
\citep[e.g.][]{Ver02}, the
relation of their numbers to mass might be non-linear, yielding
a different slope in the more extreme environments traced by the LRGs
than the slope found when averaging over all environments.

Figure \ref{fig:a125} is not quite a linear relation.  If we instead 
plot the density versus the luminosity raised to the 1.5 power, then 
we find a tight linear relation with a non-zero intercept.  This is 
shown in the first panel of Figure \ref{fig:a6}.  The horizontal axis
reads as luminosity relative to $L^*$ ($M_g^*=-20.35$), but we have
stretched the axis into a $L^{1.5}$ dependence.  We find that the fit $AL^{1.5}+B$
is a good fit in all cases, and it is slightly better (2--$\sigma$) in most cases than
$AL+B$, particularly at larger scales.  The $\chi^2$ per degree of freedom
is typically about 1, taking the jackknife errors in different luminosity
bins to be independent.
We will therefore quote our quantitative results as the parameters
of this fit.  

Figure \ref{fig:a6} shows $\phi_0V\Delta$ as a function of LRG luminosity
for 6 different scales, increasing by factors of two from $a=0.125\hmpc$ 
to $a=4\hmpc$, proper.  
We have overplotted the best $AL^{1.5}+B$ fit in each case.
The parameters of the best-fit line
for these and other subsamples are given in Table \ref{tab:fits212}
and \ref{tab:fits218}.  We quote the intercept of the linear fit
at an intermediate luminosity value chosen so that the error in the
intercept is minimized and is not covariant with the estimate of the slope.
The errors on the slopes indicate that the luminosity dependence
of clustering is detected at about 20--$\sigma$ over the range
$-22.7<M_g<-21.2$.

Figure \ref{fig:a6} and Table \ref{tab:fits212} show that there is a significant
change in slope as a function of scale.  
This is shown directly in Figure \ref{fig:a6}d, where the best-fit
lines from $a=0.125\hmpc$ and $a=4\hmpc$ have been rescaled and overplotted.
Larger scales have a softer
run of $\Delta$ against luminosity; for example, $a=4\hmpc$, the
overdensity varies
only by a factor of 2 across the same range of luminosity.  Turning this around, this means
that high luminosity LRGs have a steeper scaling of $\Delta$ with scale.
In Table \ref{tab:fits212}, we report a slope of $0.110\pm0.005$ for $a=0.125\hmpc$ and 
$0.058\pm0.010$ for $a=2\hmpc$; treating these two widely separated scales as
independent, this means that the scale dependence of
the luminosity dependence is detected at 4.6--$\sigma$.

\begin{figure}[tb]
\plotone{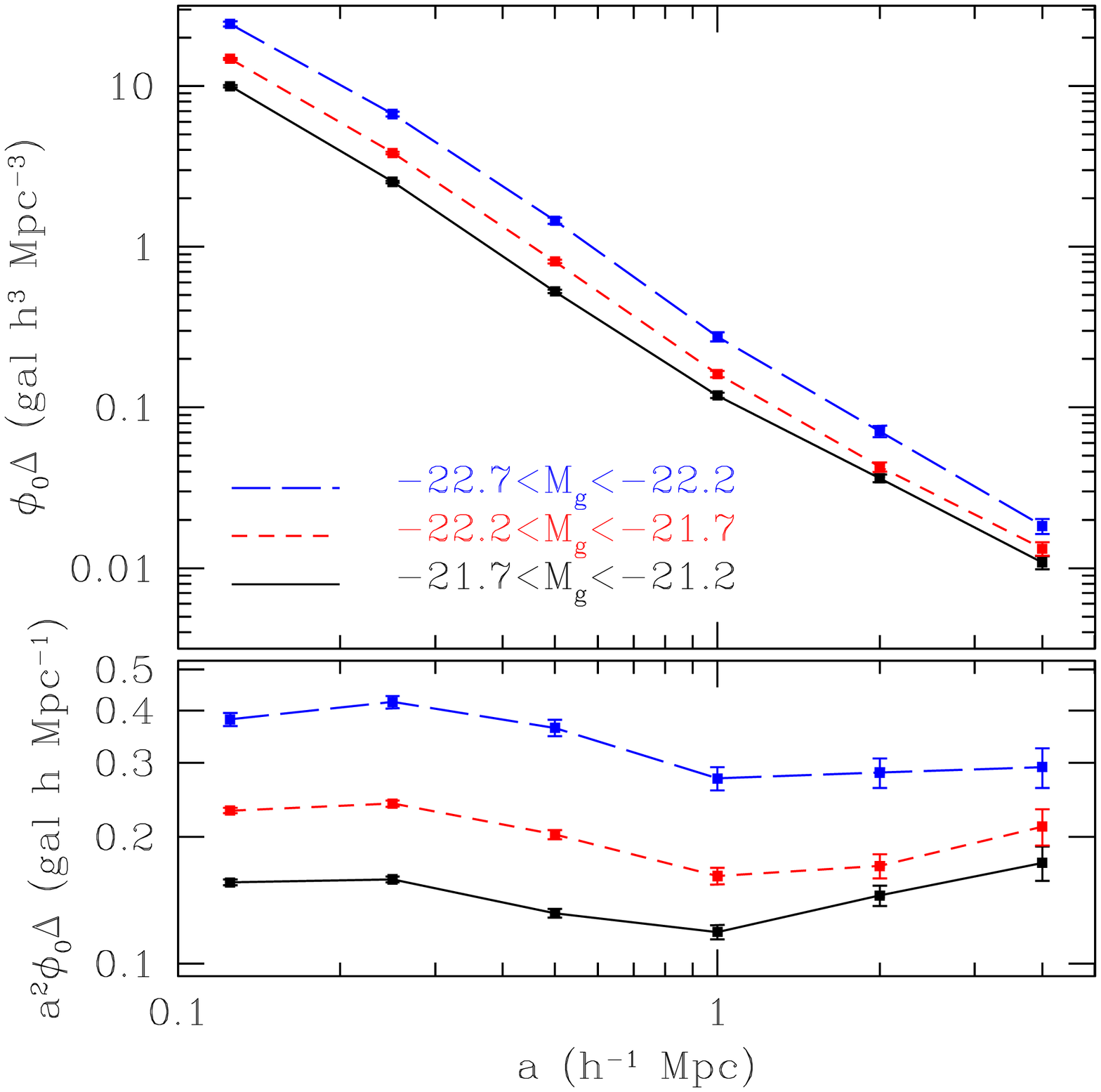}
\caption{\label{fig:r_env_Mg}%
Mean overdensity around red galaxies as a function of proper scale 
in three different luminosity bins.
The top panel shows the quantity $\phi_0\Delta$; the bottom panel
shows $a^2\phi_0\Delta$, which is chosen to flatten out the curves
so that one can see the fine detail.
The deviations from a power-law $\xiis$ are clearly visible, as is the 
luminosity dependence to the clustering.  More subtle is the 
fact that the luminosity dependence is slightly stronger at the 
smaller scales.
The reader is reminded that the $\Delta$ statistic measures
the correlation function on scales of $\sim\!1.75a$.
Only galaxies in the range $0.20<z<0.36$ have been used.  
The ``1.0'' imaging sample is used ($M^*-0.5$ to $M^*+1.0$).
}
\end{figure}

We next consider the detailed shape as a function of scale $a$.  
Figure \ref{fig:r_env_Mg} shows the value of $\phi_0\Delta$ as a function 
of $a$ for 3 coarser bins in luminosity.  The bottom panel shows
the residuals relative to an $a^{-2}$ power law.  Deviations from
the power-law model are obvious.
Using the full covariance matrices, the best-fit power laws have,
from high to low luminosity, $\chi^2 = 18$, 56, and 88 
for 4 degrees of freedom, and hence power laws are strongly rejected
with goodness-of-fit probabilities of 
$10^{-3}$, $2\times10^{-11}$, and $3\times10^{-18}$, respectively.
The deficit at $\sim1\hmpc$ and excesses at smaller and larger scales 
match the behavior
seen in the SDSS MAIN sample in \citet{Zeh04a} and in the LRG sample
in \citet{Zeh04c}.
It is interesting to interpret these correlations in the halo
occupation model as a sum of two terms, a small-scale component
in which both galaxies are in the same halo and a large-scale
component in which both galaxies are in different halos.  Clearly,
both terms would be required to fit Figure \ref{fig:r_env_Mg}.

The fact that the high and low luminosity curves in Figure \ref{fig:r_env_Mg}
are closer together at larger scale than at lower scale is another
manifestation of the mild scale dependence of the luminosity dependence.

As stated in \S\ref{sec:method}, measurements at different scales are
covariant.  We have used our 50 jackknife samples to construct the
6-dimensional covariance matrix for Figure \ref{fig:r_env_Mg} and its
``0.4'' imaging sample equivalent.  With only 50 jackknife samples, the results are noisy,
but the covariance between scales appears to reduce from unity by about
50-70\% per factor of two in scale (i.e., nearest scales are 50-70\% covariant, next
nearest are 25-49\% covariant, etc.).  The lower-luminosity samples, 
i.e. $-21.7<M_g<-21.2$, are more covariant (70\%), whereas the more luminous LRGs
are less covariant (50\%).  Widely separated scales, such as in the $0.125\hmpc$
to $2\hmpc$ comparison above, are at most 20\% covariant.

Tables \ref{tab:fits212} and \ref{tab:fits218}
show mild evidence 
for a 10\% decrease of $\phi_0\Delta$ with increasing redshift (in the non-color-selected samples).  
We see no significant evidence for a change in slope
of $\Delta$ versus $L^{1.5}$ as a function of redshift.  
As stated in \S\ref{sec:method}, both stable clustering and linear
theory predict a nearly redshift independent result.
Unfortunately, uncertainties in the redshift evolution of the imaging sample 
currently prevent us from describing the exact evolution of $\phi_0\Delta$
and confirming these hypotheses.
If the results from the extrapolation of the $z=0.1$ SDSS luminosity-color
function (Table \ref{tab:magz}) are correct, then the comoving density
of the imaging sample is increasing slightly with redshift, such that 
the $\phi_0\Delta$ for a sample of fixed comoving density would be scaling
as $(1+z)^{-2}$.  This would argue for some anomaly beyond stable clustering,
perhaps caused by some kind of unmodeled evolution.
However, it may be that cluster galaxies evolve similarly to our model and that
it is the field galaxies that are driving the evolution of the luminosity function 
\citep[e.g.][]{Lin99},
in which case stable clustering would predict constant $\phi_0\Delta$ regardless
of the bulk evolution in $\phi_0(z)$.
Calibrating our empirical measurement will require more precise modeling of the redshift
distribution of SDSS galaxies at these flux levels.
In principle, we could measure the differential evolution between different scales,
but the predicted variations are smaller than our measurement uncertainties.
On the positive side, the insensitivity to redshift validates our averaging 
$\phi_0\Delta$
over galaxies at
many redshifts; any deviations from a volume-limited LRG sample (which
are minor in any case) would produce negligible biases in $\Delta$.

\subsection{Fraction of red galaxies}

\begin{figure*}[p]
\plotone{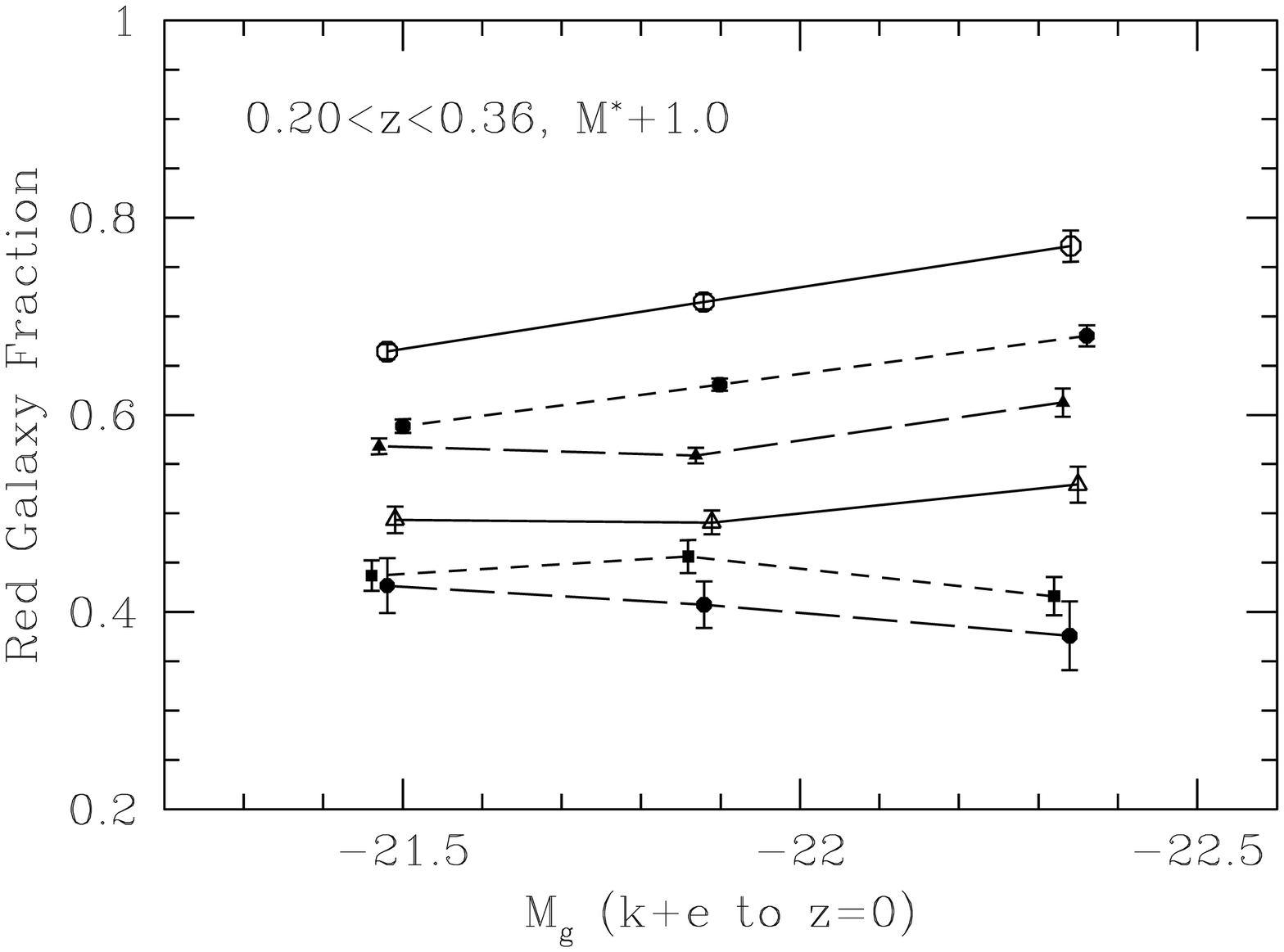}
\plotone{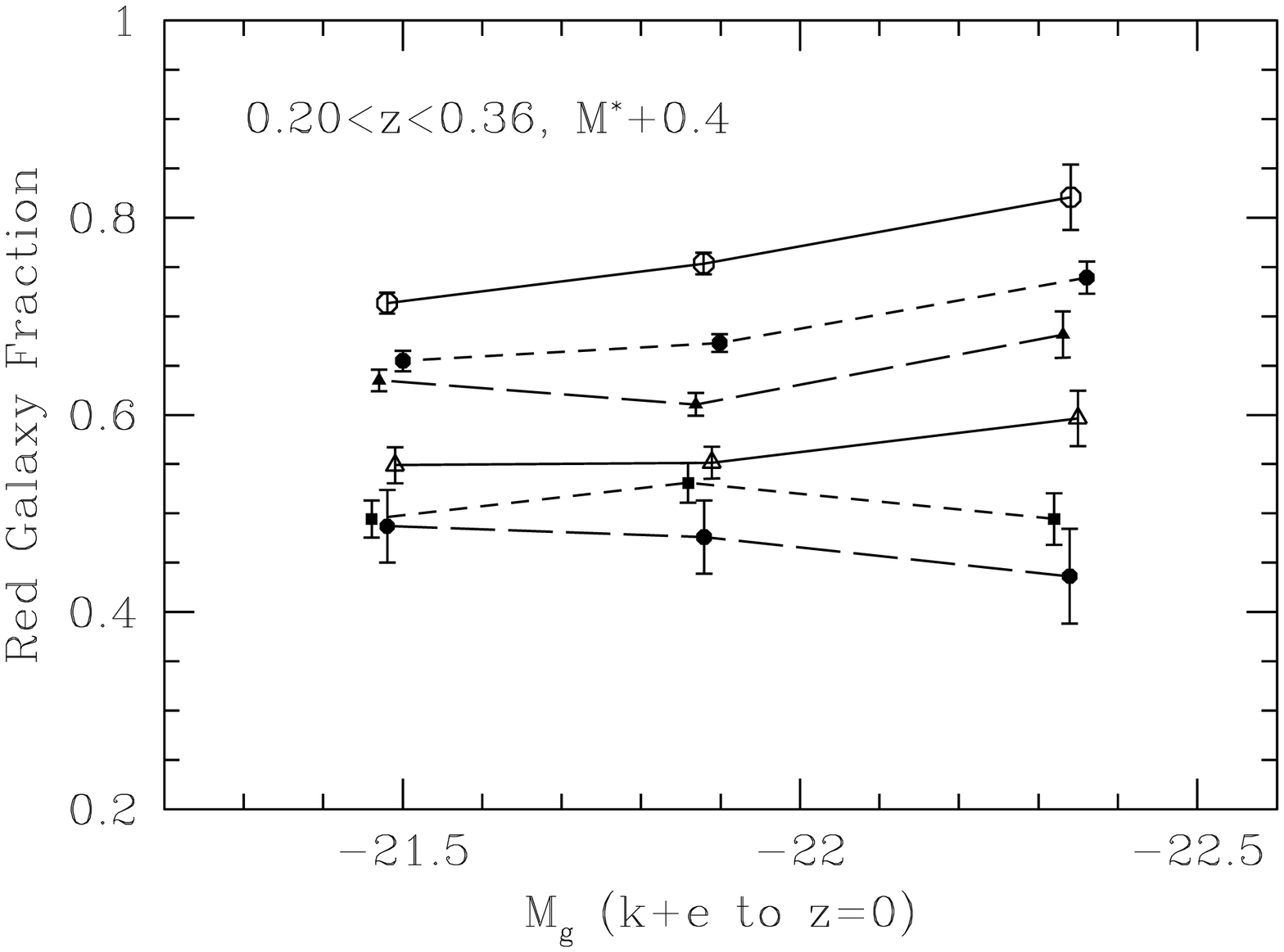} \\
\plotone{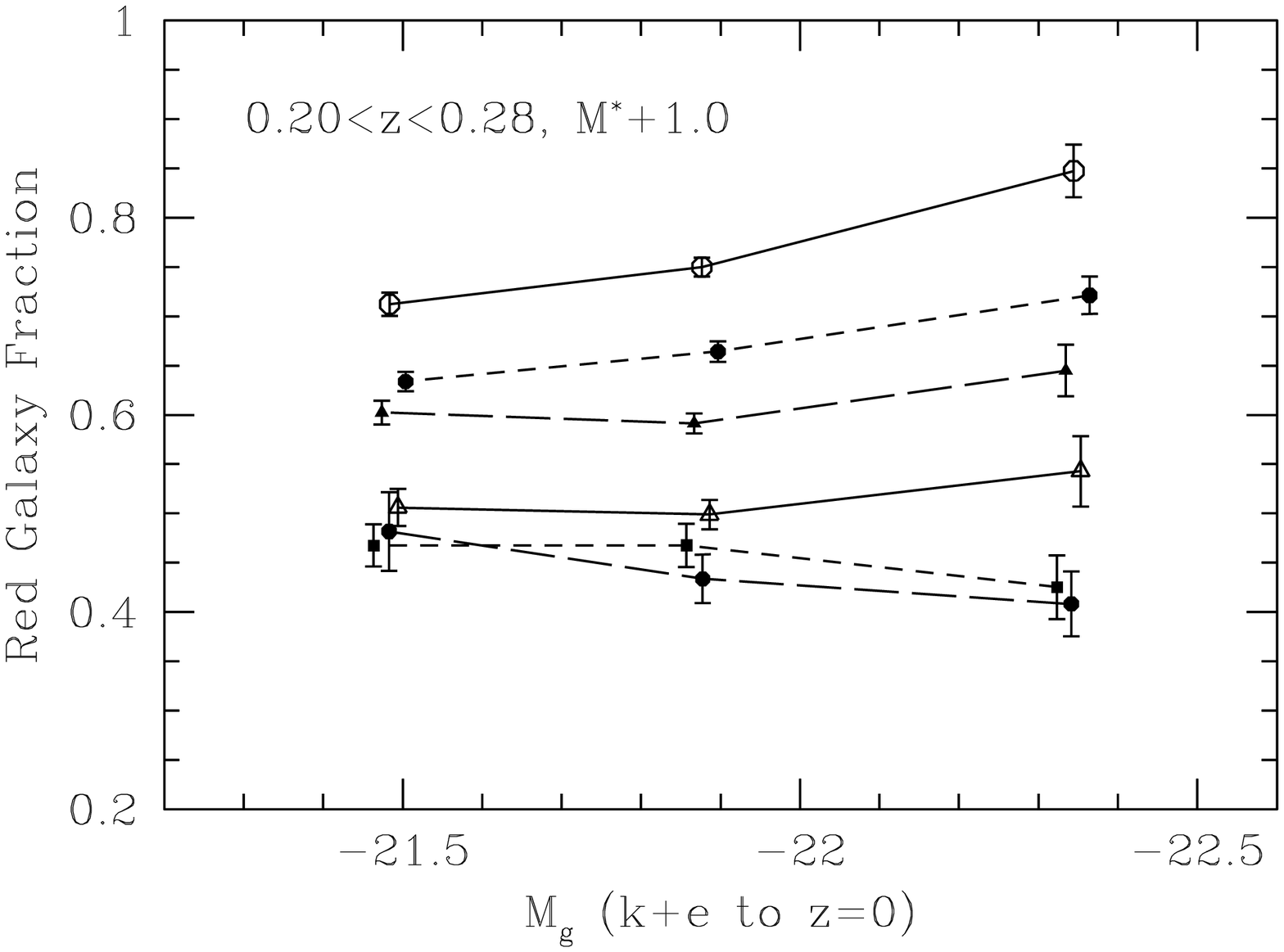}
\plotone{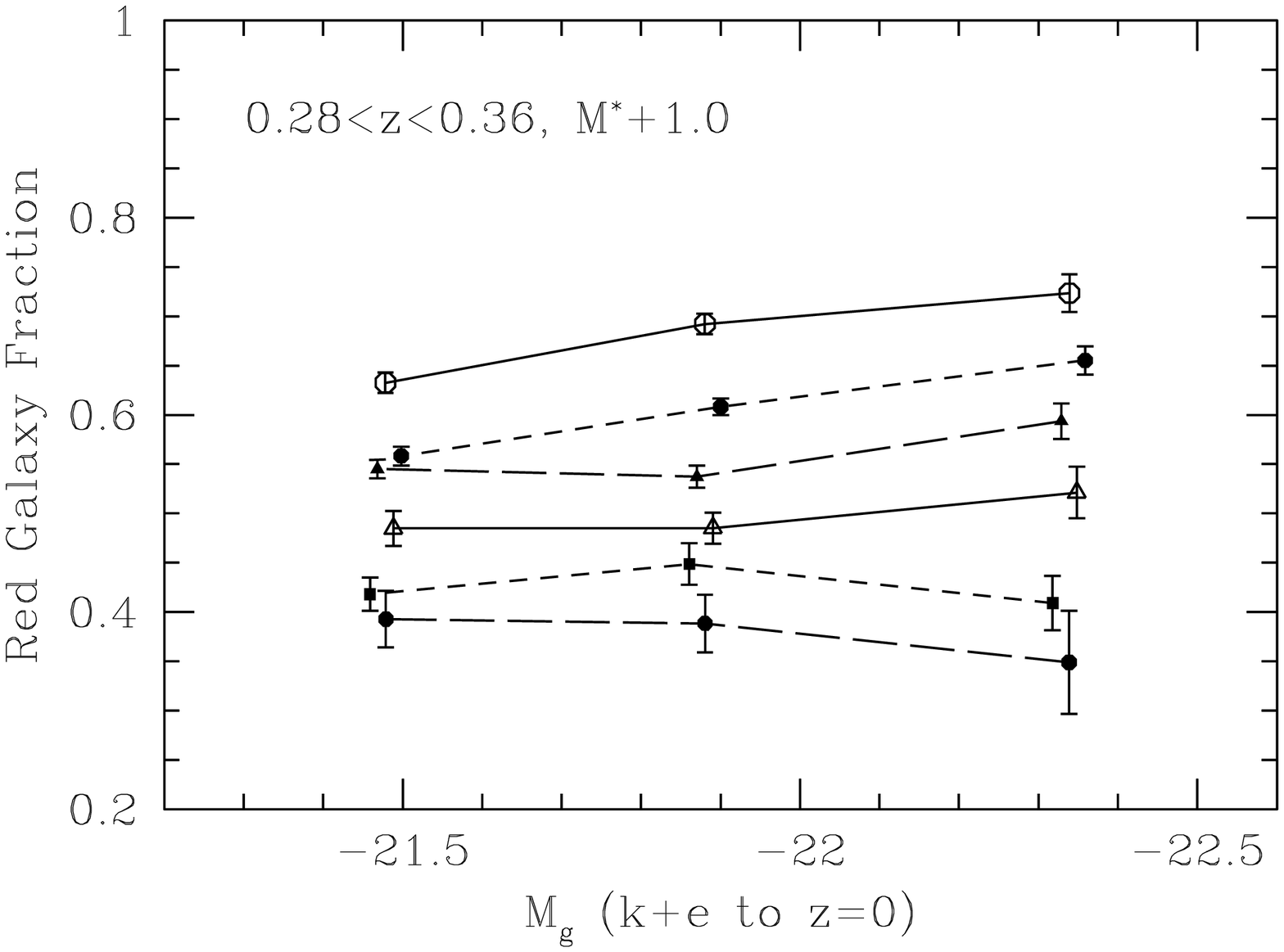} \\
\caption{\label{fig:redfrac}%
The fraction of red galaxies in the imaging sample as a function of LRG luminosity for
different scales.  Each line is a different scale, from
$a=0.125\hmpc$ to $4\hmpc$ by factor of two, from top to bottom.
A red galaxy is defined as being within $\delta(g-i)$ mag in observed
$g-i$ color from the red sequence color at the LRG redshift (see Table \protect\ref{tab:magz});
this classification is {\it not} redshift independent and so
the fraction of red galaxies is hard to quantify versus redshift.
However, because the samples are volume-limited, comparisons
between scales and LRG luminosities are well posed.
The errors are computed from the variations in the red fraction among
the jackknife samples, not from propagation of errors in $\phi_0\Delta$.
{\it (Upper left)} $M^*+1.0$ sample for $0.20<z<0.36$.
{\it (Upper right)} $M^*+0.4$ sample for $0.20<z<0.36$.
{\it (Lower left)} $M^*+1.0$ sample for $0.20<z<0.28$.
{\it (Lower right)} $M^*+0.4$ sample for $0.28<z<0.36$.
}
\end{figure*}

We define the red fraction as the 
ratio of $\phi_0\Delta$ for our red galaxy imaging sample to that for the
full imaging sample.
Because $\phi_0V\Delta+\phi_0V$ really
is the average number of galaxies, weighted by $W(r)$, surrounding our
LRGs, this ratio of these quantities is the fraction of those galaxies that are red,
as defined in \S\ref{sec:imaging}.  We neglect the homogeneous term $\phi_0V$
because it is negligible at small scales (only 10\% of the clustered term 
at $a=1\hmpc$ and dropping as $a^2$ 
below that), whereas the red fractions at larger scales are close enough to the field
values that including the homogeneous term wouldn't change the ratio appreciably
(e.g., for $a=4\hmpc$, the results might drop by 3\%).

Figure \ref{fig:redfrac} shows the red galaxy fraction as a function of scale
and LRG luminosity.
The red galaxy fraction is a strong
function of scale $a$, as one would expect from the excess of early-type galaxies
in dense regions \citep{abell65,oemler74,melnick77,dressler80,postman84}.  
We see a slight increase in the red fraction
around higher luminosity LRGs, but only on small scales.  On scales
above 1 Mpc, we do not detect a trend with luminosity.
The higher
luminosity ``0.4'' imaging sample has a slightly higher red fraction,
as one would expect from the color-magnitude distribution of galaxies.

We display two different redshift bins in the 
bottom panels of Figure \ref{fig:redfrac}.  The higher redshift
bin has a smaller red fraction.  While this trend is in the same 
direction as the conventional wisdom that galaxies are bluer at
higher redshift, it is also possible that our selection definition
has created a moving target,
as discussed in \S\ref{sec:imaging}
One effect that is certainly present is that our $r$-band flux
limits select galaxies at bluer rest-frame wavelengths at higher 
redshift, causing the samples to tilt toward a bluer fraction.
In addition, our definition of a ``red'' galaxy could be 
redshift dependent due to imperfect modeling of the rest-frame colors.  
It is interesting to note that the red fractions at the $a=4\hmpc$
scale are only slightly higher than the fractions predicted from
the low-redshift SDSS luminosity-color function in Table \ref{tab:magz}.
This suggests that at $\sim\!7\hmpc$ ($10\hmpc$ comoving), one has
nearly converged back to the field value, despite having chosen the
galaxies by their proximity to a LRG.

\begin{figure}[tb]
\plotone{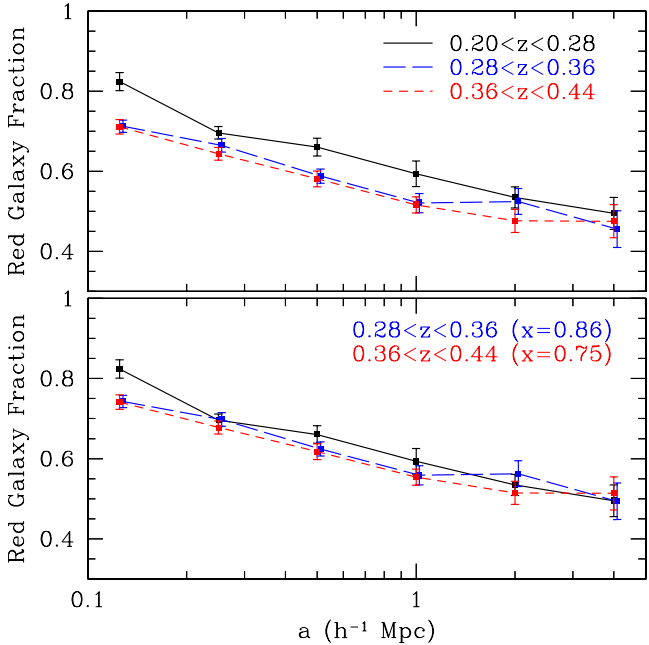}
\caption{\label{fig:scale_redfrac}%
({\it Top panel}) The fraction of red galaxies in the imaging sample as a function of 
scale for different redshift bins.  In all cases, only LRGs in
the range $-22.2<M_g<-21.8$ are used, and the imaging galaxies 
are from the ``0.4'' sample.
The points with error bars are the measurements
with $0.20<z<0.28$, $0.28<z<0.36$, and $0.36<z<0.44$, from top to bottom.  
({\it Bottom panel}) The same, but the two higher redshift bins have been rescaled
according to the model that the blue galaxies are to be diluted by 86\% and 75\%,
respectively.  These numbers were determined from extrapolation of the MAIN sample
in the absence of any evolution of intrinsic color.
The red fractions on larger scales are consistent after this rescaling; the smaller scales
show a significant difference.
}
\end{figure}

The top panel of Figure \ref{fig:scale_redfrac} shows the red fraction as a function
of scale for 3 redshift bins and the ``0.4'' imaging sample.  Again,
the red fraction is redshift dependent.
In the bottom panel, we attempt to correct for the simplest effect, namely
that lower luminosity blue galaxies can enter the sample at higher redshift
simply because of more favorable $k$-corrections.  We do this by simply
reducing the number of blue galaxies at higher redshift by some fraction $x$ 
($x=1$ being no change).  
If the red fraction at the higher redshift is $f_2$, then the fraction $f_1$ at
the lower redshift, having diluted the blue galaxy density by $x$, will
be $f_1 = (1-x+x/f_2)^{-1}$.  Note that this is not simply a multiplicative
scaling in the red fraction.  From the model in \S\ref{sec:imaging} and 
Table \ref{tab:magz} in which we extrapolate the low-redshift luminosity-color
function without any color evolution, we derive $x=0.86$ (0.75) for $0.28<z<0.36$
($0.36<z<0.44$) relative to the $0.20<z<0.28$ sample.
Applying these corrections, we see that the red fractions at large scales
(i.e., the field) are consistent with being independent of redshift.
On smaller scales, there is a trend with redshift, suggesting
that in regions near LRGs (i.e., high-density regions), galaxies have reddened
with time.  Strictly speaking, this is the evolution of a population defined
by location rather than one that is consistently tagged across time.

We stress again to the reader that variations in the definition of ``red''
as a function of redshift could still move the curves up or down,
so one should not conclude that there is no evolution of the red fraction
in the field.
These uncertainties should be amenable to better modeling of the galaxy
$k$-corrections and luminosity functions.  For example, differential
luminosity evolution between red and blue galaxies \citep{Lin99} could alter the 
red fraction predictions from Table \ref{tab:magz}.
Nevertheless, the conclusion that the evolution of the red fraction is
different between regions near LRGs and far from LRGs is robust,
as we have used the same definition of ``red'' at all scales.

We also caution against interpreting galaxies near LRGs as necessarily
representing a cluster (e.g., mass above $10^{14}\msun$) population, 
as not all LRGs live in clusters \citep[e.g.,][]{Loh03}.  However, the
densities in these regions are high on average, about $16\ihmpcC$ for 
the $a=0.125\hmpc$ case, and so our results will surely bear on the
question of differential evolution between high and low density 
environments \citep[e.g.][]{Has98,Bal99,Mar01,lewis02,gomez03,deP04}.

\section{Discussion}

We have demonstrated that the environments of LRGs, as measured by
the surrounding overdensity of $L^*$ galaxies, varies strongly with
luminosity.  Across the range from $2.2L^*$ to $8.3L^*$, the clustering
amplitude changes by a factor of 4 on $0.2\hmpc$ scales and a 
factor of 2 on $7\hmpc$ scales.  
This trend was clearly seen before \citep{norberg01,zehavi02,Hog03a}, but here we use
the large volume of the LRG sample to bring this result to very
high signal-to-noise ratio.
Moreover, the variation of the
slope versus luminosity as a function of scale implies that LRGs have 
a scale-dependent bias that
varies with luminosity.  Higher luminosity LRGs have even more
clustering on sub-Mpc scales than one would project from their
clustering on $\sim5\hmpc$ scales.

The variation of clustering amplitude as a function of LRG luminosity 
is striking in its strength, and of course these galaxies
are already more clustered than less luminous galaxies.  This clearly
points to a significant change in the masses of the host halos of LRGs
as a function of luminosity.

The clustering as a function of scale shows clear deviations from a power law,
in a manner quite consistent with the LRG autocorrelation function 
\citep{Zeh04c} and with less luminous samples \citep{Zeh04a,Zeh04b}.  
The interpretation in terms of the one-halo
and two-halo terms of the halo occupation model 
\citep{ma00,peacock00,seljak00,scoccimarro01,berlind02} would naturally
explain the qualitative structure \citep{Ber03,Mag03,Scr03,Zeh04a}; we will pursue quantitative
analyses in future papers.  Combining the cross-correlations methods employed
here with the LRG and $L^*$ galaxy auto-correlations should yield 
new constraints on the LRG and $L^*$ galaxy populations in massive halos.

The fraction of red galaxies as a function of scale and luminosity
are qualitatively consistent with the familiar density-morphology relation
\citep{dressler80,postman84}.  Higher luminosity LRGs are surrounded by a slightly larger
fraction of red galaxies at sub-Mpc scales.

Tracking redshift evolution is a challenge to any clustering method, as it
requires detailed understanding of the evolution of the selection.
In our case, this is phrased as the 
need for a measurement of the number density $\phi_0$ of the imaging sample.
Redshift evolution of differential effects can still
be measured robustly.  We find, for example, that the relation of 
red fraction versus scale changes with redshift, such that the high-density
regions near LRGs have reddened more than low-density regions far from LRGs.
This result cannot be avoided simply by rescaling the number of blue galaxies,
thereby addressing the most obvious gap in our modeling.  We intend to 
pursue this issue further in future work.
More generally, with precise modeling of the redshift distribution of the galaxies in the 
SDSS images, we could recover $\phi_0$ and measure the evolution of $\Delta$.

We propose that the scale-dependent clustering of LRGs could serve
as a test of the formation theories for these galaxies.  In particular,
the strong clustering on small scales is a challenge for theories 
that rely on passive evolution from high redshift, with no 
environment-dependent evolution of the galaxy.  If LRGs sit in
only the most massive halos, then they will be highly clustered \citep{Kai84,BBKS,Mo96,She01}, but
the correlation between their halos' mass today and their environment
at high redshift may not be sufficiently tight to allow the high-redshift 
environment to dictate today's luminosity.  Processes that
enhance the LRG luminosity in situ would loosen these constraints.

More generally, this work highlights the considerable advantage
of using cross-correlations between imaging and spectroscopic data
sets for the study of the small-scale clustering of rare types of
galaxies.  On small scales, one is nearly always shot-noise limited,
and auto-correlation analyses will suffer two portions of shot noise for
rare objects.  Cross-correlations with more populous sets of galaxies
only incur the shot noise once.  Of course, one would expect that the
clustering trends are reduced in the cross-correlation, but this is a
mild loss of signal compared to the gains in the signal-to-noise ratio.  Moreover, for
luminous galaxies, acquiring spectroscopic redshifts for the fainter $L^*$
companions is doubly challenging, as they are both faint and numerous.
Angular cross-correlations therefore allow one to leverage a considerable
amount of imaging data.  Because the correlations effectively impose a
redshift on the imaging data, one can extract rest-frame properties from
the imaging set (such as the red fractions presented here) despite not
having true redshifts.

We expect that this method could have considerable application to deep
wide-field data from the next generation of ground-based surveys or from
the Spitzer satellite, as it offers a precise and quantitative measurement
of clustering with a minimum of spectroscopic information.  

\bigskip

DJE and IZ are supported by grant AST-0098577 from the National Science
Foundation.  DJE was further supported by an Alfred P. Sloan Research Fellowship.

Funding for the creation and distribution of the SDSS Archive has
been provided by the Alfred P. Sloan Foundation, the Participating
Institutions, the National Aeronautics and Space Administration,
the National Science Foundation, the U.S. Department of Energy, the
Japanese Monbukagakusho, and the Max Planck Society. The SDSS Web site
is http://www.sdss.org/.

The SDSS is managed by the Astrophysical Research Consortium (ARC)
for the Participating Institutions. The Participating Institutions are
The University of Chicago, Fermilab, the Institute for Advanced Study,
the Japan Participation Group, The Johns Hopkins University, Los Alamos
National Laboratory, the Max-Planck-Institute for Astronomy (MPIA), the
Max-Planck-Institute for Astrophysics (MPA), New Mexico State University,
University of Pittsburgh, Princeton University, the United States Naval
Observatory, and the University of Washington.

\appendix
\section{Implementation Details}

When implementing the summation for the $\Delta_j$, 
it is easy to take account of masks and boundaries as well as to avoid
summing over very widely separated pairs.  Following \citet{Eis03},
we limit the explicit summation at some $R_{max}$ (we use $10a$)
and add the remaining piece
\beq\label{eq:DeltaErr}
\Delta_{outer} = {\nbar\over \phi_0V} \int_\Rmax^\infty 2\pi R\,dR\;[1+\wis(R)] G(R) 
\eeq
Here, $\nbar$ is the number density of objects from the imaging catalog
(selected at the chosen redshift) per unit area on the sky (measured
in transverse distance units).
The homogeneous term (the 1 in the square brackets) 
has an easy integral involving $F(R)$, whereas the correlated term can be easily
done using a power-law approximation for $\wis$ and the large-radius
expansion for $G(R)$.  If $\xiis\propto r^{-1-\alpha}$ so that 
$\wis\propto r^{-\alpha}$, then we have 
\beq\label{eq:w2xi}
{\nbar\over\phi_0}\wis(R) = R\xiis(R){\Gamma({\alpha\over2})\sqrt\pi\over 
\Gamma({\alpha+1\over2})}
\eeq
and so
\beq
\Delta_{outer} = {\nbar\over \phi_0} {2\pi F(\Rmax)\over V}
	- \xiis(\Rmax) {\Gamma({\alpha\over2}) \sqrt\pi\over \Gamma({\alpha+1\over2}) \pi(1+\alpha)}.
\eeq
The second term depends on $\xiis$, which is what we are trying to find
via $\Delta$.  However, if the correction is small, then it is a good
approximation to keep the shape of $\xiis$ fixed while trying to find
its amplitude.  In this case, the second term becomes
\beq
\Delta_{outer,clust} = 
	- \Delta {\xiis(\Rmax)\over \xiis(a)} {\beta \over \pi(1+\alpha)}.
\eeq
where 
\beq
\beta = {\Gamma({\alpha\over2}) \sqrt\pi\over \Gamma({\alpha+1\over2}) }
{3\over 2}\sqrt{\pi\over 2} \sqrt{2}^{\alpha} {1\over \Gamma(2-{\alpha\over 2})}
\eeq
$\beta=10.49$ for $\alpha=0.75$ and $\beta=3\pi$ for $\alpha=1$.

Hence, to include the effects of the upper limit to the explicit summation,
we now have
\beqa
\Delta_j &=& {1\over \phi_0V}
   \left[-2\pi \nbar F(\Rmax) + \sum_{k\in\{im\}, R_{jk}<\Rmax} G(R_{jk})\right]
   \nonumber\\ 
   &&- \Delta {\xiis(\Rmax)\over \xiis(a)} {\beta\over \pi(1+\alpha)}
\eeqa
We now would average the $\Delta_j$ over many spectroscopic objects 
to estimate $\Delta$.  However, note that the last term is just linear
in $\Delta$, so we can instead compute
\beq
\Delta_j' = {1\over \phi_0V}
   \left[-2\pi \nbar F(\Rmax) + \sum_{k\in\{im\}, R_{jk}<\Rmax} G(R_{jk})\right]
\eeq
for each object, and then estimate $\Delta$ for any subset as
\beq
\Delta = {\left<\Delta_j'\right>\over 1 + {\xiis(\Rmax)\over \xiis(a)} {\beta\over \pi(1+\alpha)}}
\eeq
where the angle brackets indicate averaging over a subset of spectroscopic objects.

An inner integration limit can be treated the same way, 
although the integral for the correlated term is different.
\beqa
\Delta_{inner} &=& {\nbar\over \phi_0V} \int_0^\Rmin 2\pi R\,dR\;[1+\wis(R)] G(R) \\
&=& 
{\nbar\over \phi_0} {2\pi F(\Rmin)\over V}
    + {2\pi\Rmin \xiis(\Rmin)\over V}
	{\Gamma({\alpha\over2}) \sqrt\pi\over \Gamma({\alpha+1\over2})}\times
   \nonumber\\ 
&&	\int_0^\Rmin R\,dR\;G(R) \left(\Rmin\over R\right)^\alpha
\eeqa
For the $W(r)$ in equation \ref{eq:Wr}, the second term becomes
\beq
\Delta_{inner,clust} = 
	\Delta \left(\Rmin\over a\right)^5 {\xiis(\Rmin)\over \xiis(a)} 
	{\beta \over \sqrt{8\pi}(4-\alpha)}.
\eeq
Because of the apodization of the window at small radii, this term is
quite small, typically of order $(\Rmin/a)^3$.
However, other choices of $W(r)$ that have $W(0)\ne 0$ will have larger
contributions.

Finally, the same principles apply to masked regions.  As described
in \citet{Eis03}, it is
easy to solve the problem by Monte Carlo by creating a dense set of 
random points on the sky {\it outside} of the survey (i.e., filling the
masks and any border regions).  We then add to each $\Delta_j$ a
quantity computed by summing over the random points found within the
$\Rmin$ to $\Rmax$ radial interval, subject to the weighting 
$G(R)[1+\wis(R)](\nbar/\nbar_{random})$,
where $\nbar_{random}$ is the areal density of the random catalog
in the same units of $\nbar$.
Again, the correlated term can be finessed by approximating 
$\wis(R) = \wis(a) (R/a)^{-\alpha}$, as it then becomes
\beq
\Delta_{mask,clust} = {a\over V\nbar_{random}} \Delta \beta 
\sum_{k\in\{ran\}} G(R_{jk}) \left(R_{jk}\over a\right)^{-\alpha}
\eeq

Hence, in practice, we compute
\beqa
\Delta_j' &=& {1\over \phi_0V}
   \left[
   \sum_{k\in\{im\}, R_{jk}<\Rmax} G(R_{jk})\right.
   \nonumber\\ 
   &&+ 2\pi \nbar (F(\Rmin)-F(\Rmax)) 
   \nonumber\\ 
   &&+ \left.{\nbar\over \nbar_{random}} \sum_{k\in\{ran\}, R_{jk}<\Rmax} G(R_{jk})
   \right]
\eeqa
and 
\beq
M_j = {a \beta\over V\nbar_{random}} 
    \sum_{k\in\{ran\}, R_{jk}<\Rmax} G(R_{jk}) \left(R_{jk}\over a\right)^{-\alpha}
\eeq
For a particular subset of spectroscopic objects, we estimate $\Delta$ as
\beq
\Delta = {\left<\Delta_j'\right> \over
    1 - \left<M_j\right> + 
    {\beta\over \pi(1+\alpha)} {\xiis(\Rmax)\over \xiis(a)}
    - {\beta\over \sqrt{8\pi}(4-\alpha)} \left(\Rmin\over a\right)^5 {\xiis(\Rmin)\over \xiis(a)} }
\eeq
Again, $\beta\approx10$ and $\alpha$ is typically 0.7 to 1.
Obviously, one would like to pick $\Rmax$ and $\Rmin$ so that 
the corrections in the denominator are minimal; we use $\Rmax = 10a$,
which should be less than 2\% correction, and $\Rmin=30h^{-1}$ kpc,
which is a tiny correction.
We exclude any objects for which the mask covered more than 40\% of
the $\Rmax$ circular region, but in practice most objects have very
little masking.  We find $\left<M_j\right>$ to be 1\% or less,
even for $a=4\hmpc$ with $\Rmax=40\hmpc$.  For $\Rmax=40\hmpc$,
about 1/3 of the objects have some fraction of their region masked;
this number drops rapidly for smaller $\Rmax$.

Our formulae use the flat-sky approximation, which is well-satisfied
since the angle on the sky even for the largest $\Rmax$ ($40\hmpc$) and closest
LRGs is only $5^\circ$ and the angles contributing significantly
to the integrals are yet smaller.  The residuals would be of order
$\theta^2$, which is $\ll1$\%.  We note that it is important to define 
the flat-sky
radius $R$ as $2D\sin(\theta/2)$ for distance $D$ to the LRG and separation 
angle $\theta$ between the two objects, as this preserves a homogeneous 
distribution from the sphere to the flat sky and
permits the unclustered background to cancel exactly from $\Delta$.

At present, our mask only includes the major boundaries of the survey,
not small-scale effects such as bright stars or bad columns.
If these masks were uncorrelated with the catalog of spectroscopic objects,
then the neglect of these regions would not bias the values of $\Delta$.
Of course, the masks and catalogs are correlated.  We assess the 
amount of spurious correlation by using an catalog of M stars as
our ``spectroscopic'' set.  We choose reasonably bright, extremely red 
M stars, as this region of color-magnitude space has essentially no 
galaxy contaminants that could correlate with the imaging catalog.
We place the stars fictitiously at $z=0.3$ and compute the correlations
with the galaxies.  The results are consistent with zero, with errors
that are roughly the same as the errors in the LRG clustering signal.
In other words, the neglect of the small-scale mask is less than a 
1--$\sigma$ effect for our measurements.

Photometric calibration errors would correlate the spectroscopic LRGs
with the imaging catalog, since both samples impose a flux limit, and
create a false signal.  However,
the effect is small.  Following equation (\ref{eq:DeltaErr}), we have
\begin{equation}\label{eq:cal}
\phi_0V\Delta_{\rm cal} = \nbar\int_0^\infty 2\pi R\;dR\;\wis G(R),
\end{equation}
where $\wis$ is now the cross-correlation between the two samples
due to the calibration error.  If $\wis$ were scale-independent for
scales near $a$, as would be the case if the calibration were wrong
in patches whose size was much larger than $a$, then this integral
would cancel to zero.  To avoid cancellation, one must have structure
in $\wis$ near the scale $a$.  The function $G(R)$ peaks at 0.8 at
$R\approx a$.  Conservatively, one would have $\phi_0V\Delta_{\rm cal}
= \nbar\pi a^2 \wis(a)$.  $\nbar\approx 8h^2 \mpc^{-2}$ for the ``1.0''
sample at $z=0.3$, so for $a=1\hmpc$, we have $\phi_0V\Delta_{\rm cal}
\approx 25\wis$.  The observed $\phi_0V\Delta$ exceeds 5 with 3\% errors.
Hence, we need the calibration-induced $\wis$ to be less than 0.006.
The SDSS rms error in the $r$ band are about 2\%, which produces a 6\%
response in the LRGs \citep{Eis01} and a 2\% response in the imaging
sample.  This implies $\wis=0.0012$, a factor of 5 smaller than required.
This is conservative because the actual correlation function from the
SDSS calibration errors is a smooth function of scale, i.e., the 2\%
errors accrue from errors on many patch sizes, which tends to cause
the integral in equation (\ref{eq:cal}) to be smaller.  Hence, angular
correlations between the samples due to calibration errors are negligibly
small for our purpose.

It is worth noting that the deprojection formalism can be applied trivially
to spectroscopic auto-correlation applications with
the usual projected correlation function
\begin{equation}
w_p(R) \equiv 2\int_{-\infty}^{\infty} dZ\;\xi_{red}(\sqrt{Z^2+R^2}),
\end{equation}
where $\xi_{red}(r)$ is the redshift space correlation function.
Defining $\Delta$ as in Equation (\ref{eq:Deltadef}), we have
\begin{equation}
\Delta = {1\over V}\int_0^\infty 2\pi R\;dR\; w_p(R) G(R)
\end{equation}
Hence, to compute $\Delta$ from an auto-correlation study, there is
no need to deproject $w_p$ to $\xi_{real}$ and then integrate to $\Delta$.

\section{Weak Lensing Effects}

We argue here that weak lensing magnification effects are small for our
analysis, only a $\sim\!1\%$ effect, comparable to our quoted errors in
the best cases.
The root reasons are simple: our flux limits select $L^*$
galaxies at the redshift of the LRG, so the luminosity implied for 
a target at a redshift high enough to be lensed effectively is much
higher, resulting in a low density of potential sources.  In more 
observational terms, at $z=0.3$ we are using $r=20$ galaxies and there
are rather few $r=20$ galaxies at $z\gtrsim0.6$ on the sky.  Furthermore,
the scales we are studying ($>0.2\hmpc$) are large compared to the 
Einstein radius of an LRG or even a reasonable cluster.

Mathematically, we consider magnification patterns of the form $(R/R_E)^{-1}$,
where $R_E$ is a constant.
This would generate an angular correlation proportional to $R^{-1}$, with
the amplitude depending on the cosmological distances and the slope of the
luminosity function.

For example, summing over source redshifts, one could write the contribution
to $\Delta$ as
\begin{eqnarray}
\phi_0\Delta_{WL} &=& {1\over V}\int 2\pi R\,dR\; G(R) 
\int_{z_{LRG}}^\infty dz\; \phi(z) \nonumber\\
&&\times \left[D_M(z)\over D_M(z_{LRG})\right]^2 (\alpha-1) \left(R_E(z)\over R\right)
\end{eqnarray}
where $z$ is the distance along the line of sight (not redshift),
$R$ is the transverse distance across the line of sight,
$D_M(z)$ is the cosmological proper motion distance,
$\phi(z)$ is the comoving density of galaxies passing the selection at the LRG redshift,
$\alpha$ is the logarithmic slope of the cumulative luminosity function,
and $R_E(z)$ is the Einstein radius (in distance, not angle) for that source redshift.
The ratio of $D_M$ enters because of the volume per solid angle,
and the $\alpha-1$ is the familiar competition between magnification of sources and
dilution of source density \citep{Tur84}.

The equation can be simplified by noting that equations (\ref{eq:xi2Delta}), (\ref{eq:w2xi}),
and 
\begin{equation}
\Delta = {\nbar\over \phi_0V}\int_0^\infty 2\pi R\,dR\; \wis(R) G(R)
\end{equation}
[see eq.~(\ref{eq:DeltaErr})] imply that 
\begin{equation}
{1\over V} \int 2\pi R\,dR\;G(R) \left(R_0\over R\right) = {R_0\over 3\pi a^2}
\end{equation}
for the $W(r)$ in equation (\ref{eq:Wr}).  This means that the weak lensing
contribution to $\Delta$ is
\begin{equation}
\Delta_{WL} = {1\over 3\pi a^2} \int_{z_{LRG}}^\infty dz\;{\phi(z)\over \phi_0}
\left[D_M(z)\over D_M(z_{LRG})\right]^2 (\alpha-1) R_E(z).
\end{equation}

One could do a careful integration using a luminosity function, but here we will
merely estimate.  $R_E$ is small unless the sources are well behind the lens.
Characteristically, we would have $z\approx 2z_{LRG}$ and $D_M(z) \approx 2D_M(z_{LRG})$.
The luminosity threshold will rise as the square of the luminosity distance,
which is $(1+z)D_M$, in addition to any $k$ corrections, so we would typically have 
the luminosity threshold increased by about a factor of 5--10.  This suggests
$\phi(z) \approx 0.01 \phi_0$.  We assume $\alpha-1\approx1$.  The line of sight 
interval would characteristically be $1000\hmpc$.  We conservatively take $R_E=20$ kpc, 
which is already cluster sized ($6''$ radius) and would correspond to 10\% magnification
ratios at $1'$ \citep[e.g.,][]{Bar01,Ben01,Jai03}.  
Putting these together suggests $\Delta_{WL} \sim 0.1 \Mpc^2/a^2$,
whereas our measured $\Delta$ values are 100 times larger.  

The calculation for magnification of the LRG by foreground galaxies is analogous
and gives a similar estimate.  Here the problem is that the foreground
galaxies are well below $L^*$ and hence are poor lenses.  The volume and path
lengths are also less favorable.

We therefore estimate the weak lensing effects at 1-2\%.
While not important for this work,
weak lensing could be an issue for some applications of the cross-correlation method,
particularly for studies involving cross-correlation to sub-$L^*$ galaxies or
to objects with favorable $k$-corrections at high redshifts (e.g.~sub-millimeter
galaxies).

{}

\clearpage

\begin{table*}[p]\footnotesize
\caption{\label{tab:fits212}}
\begin{center}
{\sc Fits to Overdensities versus Luminosity\\}
\begin{tabular}{cccccccc} 
\doubleline
$a$ ($h^{-1}$ Mpc) & Redshift$^a$ & $L_{back}$$^b$ & Color$^c$ 
& $L_{LRG}$$^d$ & $\phi_0V\Delta(L_{ref})$$^e$ 
& Slope$^f$ & $\chi^2$$^g$ \\
\doubleline
\input{environ_tab2a.tex}
\doubleline
\end{tabular}
\end{center}
NOTES.---%
$^a$ The redshift range of the LRG sample.
\\
$^b$ The minimum luminosity of the imaging galaxies in magnitudes below $M^*$.
The maximum luminosity is always $M^*-0.5$.  This table presents the ``1.0'' sample.
\\
$^c$ Marked with $g-i$ if imaging galaxies have been required to be within
$\delta(g-i)$ mag in {\it observed} $g-i$ color of the red sequence at the LRG redshift.  
\\
$^d$ The minimum $M_g$ of the LRGs used in the fit.
\\
$^e$ The value of the best-fit linear relation of $\phi_0V\Delta$ versus
$(L/L^*)^{1.5}$, evaluated at $3.3L^*$ for the $M_g<-21.2$ sample and $4.5L^*$
for the $M_g<-21.8$ sample.  These values were picked so that the errors
on the slope and intercept of the best-fit line are nearly independent.
The errors on the last digits are given in parenthesis.
We use $M_g^* = -20.35$.
\\
$^f$ The slope of the best-fit line and its error, both divided by the 
value in the previous column to give a reasonable normalization 
(but one that is different between this table and Table \protect\ref{tab:fits218}).
\\
$^g$ $\chi^2$ of the data with respect to the best-fit line.
Samples with $M_g<-21.2$ have 15 luminosity bins and hence 13 degrees of freedom.
Samples with $M_g<-21.8$ have 9 luminosity bins and hence 7 degrees of freedom.
\end{table*}

\begin{table*}[p]\footnotesize
\caption{\label{tab:fits218}}
\begin{center}
{\sc Fits to Overdensities versus Luminosity\\}
\begin{tabular}{cccccccc} 
\doubleline
$a$ ($h^{-1}$ Mpc) & Redshift & $L_{back}$ & Color & $L_{LRG}$ & $\phi_0V\Delta(L_{ref})$ 
& Slope & $\chi^2$ \\
\doubleline
\input{environ_tab2b.tex}
\doubleline
\end{tabular}
\end{center}
NOTES.---%
As Table \protect\ref{tab:fits212}, but for the higher luminosity ``0.4'' imaging
sample, which allows us to use higher redshift LRGs if we restrict to
high luminosity LRGs.
\end{table*}

\end{document}

%% file: environ_tab3.dat
0.16 & 17.889 & 0.123 &  9.17 & 41\% & 4.04 & 45\% \\
0.18 & 18.180 & 0.134 &  9.37 & 40\% & 4.14 & 45\% \\
0.20 & 18.441 & 0.146 &  9.52 & 40\% & 4.21 & 44\% \\
0.22 & 18.674 & 0.159 &  9.71 & 39\% & 4.30 & 43\% \\
0.24 & 18.895 & 0.170 &  9.92 & 38\% & 4.40 & 43\% \\
0.26 & 19.099 & 0.181 &  10.08 & 37\% & 4.48 & 42\% \\
0.28 & 19.292 & 0.193 &  10.31 & 37\% & 4.59 & 41\% \\
0.30 & 19.472 & 0.208 &  10.48 & 36\% & 4.68 & 40\% \\
0.32 & 19.638 & 0.223 &  10.66 & 34\% & 4.77 & 39\% \\
0.34 & 19.803 & 0.236 &  10.92 & 33\% & 4.90 & 38\% \\
0.36 & 19.962 & 0.245 &  11.07 & 33\% & 4.97 & 38\% \\
0.38 & 20.116 & 0.251 &  \nodata & \nodata & 5.08 & 37\% \\
0.40 & 20.269 & 0.259 &  \nodata & \nodata & 5.21 & 36\% \\
0.42 & 20.423 & 0.270 &  \nodata & \nodata & 5.36 & 34\% \\
0.44 & 20.576 & 0.285 &  \nodata & \nodata & 5.51 & 33\% \\

%% file: environ_tab2a.tex
0.125 & $0.20<z<0.36$ & 1.0 &       & $-21.2$ & 1.082(13) & 0.110(5) &  6.8 
\\
0.125 & $0.20<z<0.28$ & 1.0 &       & $-21.2$ & 1.100(21) & 0.120(8) & 10.2 
\\
0.125 & $0.28<z<0.36$ & 1.0 &       & $-21.2$ & 1.073(16) & 0.108(6) & 15.0 
\\
0.125 & $0.20<z<0.36$ & 1.0 & $g-i$ & $-21.2$ & 0.742(9) & 0.132(5) &  6.7 
\\
0.125 & $0.20<z<0.28$ & 1.0 & $g-i$ & $-21.2$ & 0.803(14) & 0.145(8) &  7.9 
\\
0.125 & $0.28<z<0.36$ & 1.0 & $g-i$ & $-21.2$ & 0.704(10) & 0.125(6) & 10.5 
\\
\singleline
0.25  & $0.20<z<0.36$ & 1.0 &       & $-21.2$ & 2.214(26) & 0.118(5) &  7.8 
\\
0.25  & $0.20<z<0.28$ & 1.0 &       & $-21.2$ & 2.287(39) & 0.131(8) &  4.4 
\\
0.25  & $0.28<z<0.36$ & 1.0 &       & $-21.2$ & 2.159(34) & 0.111(6) &  6.8 
\\
0.25  & $0.20<z<0.36$ & 1.0 & $g-i$ & $-21.2$ & 1.336(16) & 0.141(5) & 11.4 
\\
0.25  & $0.20<z<0.28$ & 1.0 & $g-i$ & $-21.2$ & 1.482(25) & 0.149(8) &  7.1 
\\
0.25  & $0.28<z<0.36$ & 1.0 & $g-i$ & $-21.2$ & 1.248(20) & 0.137(7) & 12.2 
\\
\singleline
0.5   & $0.20<z<0.36$ & 1.0 &       & $-21.2$ & 3.64(6) & 0.130(7) & 13.2 
\\
0.5   & $0.20<z<0.28$ & 1.0 &       & $-21.2$ & 3.77(8) & 0.137(10) & 10.9 
\\
0.5   & $0.28<z<0.36$ & 1.0 &       & $-21.2$ & 3.56(8) & 0.126(9) & 12.9 
\\
0.5   & $0.20<z<0.36$ & 1.0 & $g-i$ & $-21.2$ & 2.046(30) & 0.133(7) & 24.8 
\\
0.5   & $0.20<z<0.28$ & 1.0 & $g-i$ & $-21.2$ & 2.25(5) & 0.140(10) & 11.0 
\\
0.5   & $0.28<z<0.36$ & 1.0 & $g-i$ & $-21.2$ & 1.914(39) & 0.131(9) & 17.1 
\\
\singleline
1.0   & $0.20<z<0.36$ & 1.0 &       & $-21.2$ & 6.42(15) & 0.090(9) &  7.0 
\\
1.0   & $0.20<z<0.28$ & 1.0 &       & $-21.2$ & 6.72(22) & 0.089(12) & 11.1 
\\
1.0   & $0.28<z<0.36$ & 1.0 &       & $-21.2$ & 6.22(18) & 0.090(12) &  7.0 
\\
1.0   & $0.20<z<0.36$ & 1.0 & $g-i$ & $-21.2$ & 3.17(7) & 0.099(8) & 11.0 
\\
1.0   & $0.20<z<0.28$ & 1.0 & $g-i$ & $-21.2$ & 3.39(10) & 0.106(11) &  5.0 
\\
1.0   & $0.28<z<0.36$ & 1.0 & $g-i$ & $-21.2$ & 3.03(8) & 0.093(11) &  9.5 
\\
\singleline
2.0   & $0.20<z<0.36$ & 1.0 &       & $-21.2$ & 15.1(4) & 0.058(10) & 14.1 
\\
2.0   & $0.20<z<0.28$ & 1.0 &       & $-21.2$ & 15.1(6) & 0.084(14) &  7.5 
\\
2.0   & $0.28<z<0.36$ & 1.0 &       & $-21.2$ & 15.2(6) & 0.046(13) & 13.7 
\\
2.0   & $0.20<z<0.36$ & 1.0 & $g-i$ & $-21.2$ & 6.64(14) & 0.059(8) & 10.4 
\\
2.0   & $0.20<z<0.28$ & 1.0 & $g-i$ & $-21.2$ & 7.03(22) & 0.068(12) & 11.2 
\\
2.0   & $0.28<z<0.36$ & 1.0 & $g-i$ & $-21.2$ & 6.45(18) & 0.053(10) &  6.5 
\\
\singleline
4.0   & $0.20<z<0.36$ & 1.0 &       & $-21.2$ & 36.4(15) & 0.047(13) &  8.0 
\\
4.0   & $0.20<z<0.28$ & 1.0 &       & $-21.2$ & 37.4(20) & 0.082(18) & 10.1 
\\
4.0   & $0.28<z<0.36$ & 1.0 &       & $-21.2$ & 35.7(18) & 0.024(16) &  7.7 
\\
4.0   & $0.20<z<0.36$ & 1.0 & $g-i$ & $-21.2$ & 15.3(4) & 0.041(9) &  7.3 
\\
4.0   & $0.20<z<0.28$ & 1.0 & $g-i$ & $-21.2$ & 17.4(6) & 0.061(12) & 12.2 
\\
4.0   & $0.28<z<0.36$ & 1.0 & $g-i$ & $-21.2$ & 14.0(5) & 0.022(12) &  6.6 
\\

%% file: environ_tab2b.tex
0.125 & $0.20<z<0.44$ & 0.4 &       & $-21.8$ & 0.637(13) & 0.065(7) &  7.9 
\\
0.125 & $0.20<z<0.28$ & 0.4 &       & $-21.8$ & 0.645(27) & 0.085(18) &  2.7 
\\
0.125 & $0.28<z<0.36$ & 0.4 &       & $-21.8$ & 0.673(20) & 0.063(12) &  7.2 
\\
0.125 & $0.36<z<0.44$ & 0.4 &       & $-21.8$ & 0.616(20) & 0.068(11) &  5.1 
\\
0.125 & $0.20<z<0.44$ & 0.4 & $g-i$ & $-21.8$ & 0.468(8) & 0.071(7) &  4.7 
\\
0.125 & $0.20<z<0.28$ & 0.4 & $g-i$ & $-21.8$ & 0.530(21) & 0.093(17) &  3.3 
\\
0.125 & $0.28<z<0.36$ & 0.4 & $g-i$ & $-21.8$ & 0.481(13) & 0.072(11) &  2.9 
\\
0.125 & $0.36<z<0.44$ & 0.4 & $g-i$ & $-21.8$ & 0.433(11) & 0.068(10) &  2.5 
\\
\singleline
0.25  & $0.20<z<0.44$ & 0.4 &       & $-21.8$ & 1.314(23) & 0.084(7) &  6.0 
\\
0.25  & $0.20<z<0.28$ & 0.4 &       & $-21.8$ & 1.45(5) & 0.091(15) &  3.8 
\\
0.25  & $0.28<z<0.36$ & 0.4 &       & $-21.8$ & 1.355(38) & 0.067(11) &  3.5 
\\
0.25  & $0.36<z<0.44$ & 0.4 &       & $-21.8$ & 1.219(35) & 0.098(11) &  3.3 
\\
0.25  & $0.20<z<0.44$ & 0.4 & $g-i$ & $-21.8$ & 0.870(15) & 0.090(7) &  7.9 
\\
0.25  & $0.20<z<0.28$ & 0.4 & $g-i$ & $-21.8$ & 1.013(35) & 0.103(15) &  3.4 
\\
0.25  & $0.28<z<0.36$ & 0.4 & $g-i$ & $-21.8$ & 0.911(24) & 0.088(10) &  5.6 
\\
0.25  & $0.36<z<0.44$ & 0.4 & $g-i$ & $-21.8$ & 0.782(21) & 0.084(10) &  7.2 
\\
\singleline
0.5   & $0.20<z<0.44$ & 0.4 &       & $-21.8$ & 2.21(5) & 0.083(8) &  4.8 
\\
0.5   & $0.20<z<0.28$ & 0.4 &       & $-21.8$ & 2.47(9) & 0.092(17) &  0.8 
\\
0.5   & $0.28<z<0.36$ & 0.4 &       & $-21.8$ & 2.37(8) & 0.075(14) &  5.2 
\\
0.5   & $0.36<z<0.44$ & 0.4 &       & $-21.8$ & 2.00(8) & 0.091(14) &  5.3 
\\
0.5   & $0.20<z<0.44$ & 0.4 & $g-i$ & $-21.8$ & 1.342(28) & 0.098(8) & 10.4 
\\
0.5   & $0.20<z<0.28$ & 0.4 & $g-i$ & $-21.8$ & 1.63(6) & 0.106(17) &  3.1 
\\
0.5   & $0.28<z<0.36$ & 0.4 & $g-i$ & $-21.8$ & 1.41(5) & 0.103(13) & 10.0 
\\
0.5   & $0.36<z<0.44$ & 0.4 & $g-i$ & $-21.8$ & 1.18(4) & 0.097(11) & 10.7 
\\
\singleline
1.0   & $0.20<z<0.44$ & 0.4 &       & $-21.8$ & 3.68(12) & 0.067(11) & 12.2 
\\
1.0   & $0.20<z<0.28$ & 0.4 &       & $-21.8$ & 4.00(25) & 0.075(21) &  4.9 
\\
1.0   & $0.28<z<0.36$ & 0.4 &       & $-21.8$ & 3.92(19) & 0.080(20) &  4.6 
\\
1.0   & $0.36<z<0.44$ & 0.4 &       & $-21.8$ & 3.39(18) & 0.043(18) & 12.1 
\\
1.0   & $0.20<z<0.44$ & 0.4 & $g-i$ & $-21.8$ & 1.98(5) & 0.068(9) &  8.2 
\\
1.0   & $0.20<z<0.28$ & 0.4 & $g-i$ & $-21.8$ & 2.34(11) & 0.076(19) &  1.8 
\\
1.0   & $0.28<z<0.36$ & 0.4 & $g-i$ & $-21.8$ & 2.09(9) & 0.098(19) &  3.4 
\\
1.0   & $0.36<z<0.44$ & 0.4 & $g-i$ & $-21.8$ & 1.78(7) & 0.044(12) &  4.2 
\\
\singleline
2.0   & $0.20<z<0.44$ & 0.4 &       & $-21.8$ & 8.19(33) & 0.064(12) & 11.9 
\\
2.0   & $0.20<z<0.28$ & 0.4 &       & $-21.8$ &  8.7(6) & 0.078(24) &  6.9 
\\
2.0   & $0.28<z<0.36$ & 0.4 &       & $-21.8$ &  7.9(5) & 0.076(26) &  9.2 
\\
2.0   & $0.36<z<0.44$ & 0.4 &       & $-21.8$ &  8.1(5) & 0.056(20) &  8.2 
\\
2.0   & $0.20<z<0.44$ & 0.4 & $g-i$ & $-21.8$ & 4.07(11) & 0.055(9) & 10.1 
\\
2.0   & $0.20<z<0.28$ & 0.4 & $g-i$ & $-21.8$ & 4.57(24) & 0.069(21) &  5.8 
\\
2.0   & $0.28<z<0.36$ & 0.4 & $g-i$ & $-21.8$ & 4.19(19) & 0.062(18) &  4.2 
\\
2.0   & $0.36<z<0.44$ & 0.4 & $g-i$ & $-21.8$ & 3.77(16) & 0.040(14) &  4.2 
\\
\singleline
4.0   & $0.20<z<0.44$ & 0.4 &       & $-21.8$ & 20.2(13) & 0.031(17) &  5.2 
\\
4.0   & $0.20<z<0.28$ & 0.4 &       & $-21.8$ & 24.3(19) & 0.061(27) &  3.5 
\\
4.0   & $0.28<z<0.36$ & 0.4 &       & $-21.8$ & 20.5(20) & -0.003(28) &  5.0 
\\
4.0   & $0.36<z<0.44$ & 0.4 &       & $-21.8$ & 18.4(17) & 0.035(27) &  6.6 
\\
4.0   & $0.20<z<0.44$ & 0.4 & $g-i$ & $-21.8$ & 9.29(35) & 0.027(11) &  5.4 
\\
4.0   & $0.20<z<0.28$ & 0.4 & $g-i$ & $-21.8$ & 12.0(7) & 0.055(20) &  3.4 
\\
4.0   & $0.28<z<0.36$ & 0.4 & $g-i$ & $-21.8$ &  9.0(5) & 0.009(19) &  3.8 
\\
4.0   & $0.36<z<0.44$ & 0.4 & $g-i$ & $-21.8$ &  8.5(5) & 0.021(18) &  5.8 
\\